\documentclass[12pt]{article}
\pdfoutput=1

\usepackage{putex}
\usepackage{graphicx}
\usepackage{caption}
\usepackage{amsmath}
\usepackage{amssymb}
\usepackage{array}
\usepackage{multirow}
\usepackage{mathtools}
\usepackage{comment}
\usepackage{subcaption}
\usepackage{epstopdf}
\usepackage{enumerate}
\usepackage{cite}
\usepackage{youngtab}
\usepackage{tensor}
\usepackage{slashed}
\usepackage[aligntableaux=center]{ytableau}
\usepackage[utf8]{inputenc}
\usepackage{rotating}
\usepackage{bigfoot}
\usepackage[
      colorlinks=true,
      linkcolor=blue,
      urlcolor=blue,
      filecolor=black,
      citecolor=red,
      linktocpage=true
      ]{hyperref}

\newcommand{\abs}[1]{\left\lvert #1 \right\rvert}

\newcommand {\be} {\begin {equation}}
\newcommand {\ee} {\end {equation}}

\newcommand {\bes} {\begin {equation*}}
\newcommand {\ees} {\end {equation*}}

\newcommand{\es}[2] {\begin{equation} \label{#1} \begin{split} #2 \end{split} \end{equation}}

\newcommand{\cA}{{\mathcal A}}

\newcommand{\cE}{{\mathcal E}}
\newcommand{\cF}{{\mathcal F}}
\newcommand{\cG}{{\mathcal G}}

\newcommand{\cN}{{\mathcal N}}
\newcommand{\cO}{{\mathcal O}}

\newcommand{\cS}{{\mathcal S}}
\newcommand{\cT}{{\mathcal T}}

\newcommand   \zb  {\bar{z}}

\newcommand   \ab  {\bar{\alpha}}
\renewcommand \l  {\lambda}

\newcommand{\bea}{\begin{equation}\begin{aligned}}
\newcommand{\eea}[1]{\label{#1}\end{aligned}\end{equation}}

\newcommand{\beq}{\begin{equation}}
\newcommand{\eeq}{\end{equation}}

\newcommand\al{{\alpha}}

\newcommand\de{{\ensuremath{{\delta}}}}
\newcommand\De{{\ensuremath{{\Delta}}}}

\def\ie{\begin{equation}\begin{aligned}}
\def\fe{\end{aligned}\end{equation}}

\newcommand{\B}{{\beta}}

\def\ula{{\underline{\smash \lambda}}}

\def\half{{\scriptstyle \frac 12}}

\def\sevenh{{\scriptstyle \frac 72}}
\def\threeh{{\scriptstyle \frac 32}}
\def\fiveh{{\scriptstyle \frac 52}}
\def\nineh{{\scriptstyle \frac 92}}
\def\elevenh{{\scriptstyle \frac {11}{2}}}

\newcommand{\orb}{\mathfrak{o}}

\numberwithin{equation}{section}

\def\<{\langle}
\def\>{\rangle}

\begin{document}

\preprint{}

\institution{oxford}{Mathematical Institute, University of Oxford,
Woodstock Road, Oxford, OX2 6GG, UK}
\institution{Exile}{Department of Particle Physics and Astrophysics, Weizmann Institute of Science, Rehovot, Israel}

\title{Modular invariant holographic correlators for $\mathcal{N}=4$ SYM with general gauge group}

\authors{
Luis F. Alday\worksat{\oxford}\footnote{e-mail: {\tt alday@maths.ox.ac.uk}},
Shai M. Chester\worksat{\Exile}\footnote{e-mail: {\tt iahs81@gmail.com}} and
Tobias Hansen\worksat{\oxford}\footnote{e-mail: {\tt tobias.hansen@maths.ox.ac.uk}}}

\abstract{
We study the stress tensor four-point function for $\mathcal{N}=4$ SYM with gauge group $G=SU(N)$, $SO(2N+1)$, $SO(2N)$ or $USp(2N)$ at large $N$. When $G=SU(N)$, the theory is dual to type IIB string theory on $AdS_5\times S^5$ with complexified string coupling $\tau_s$, while for the other cases it is dual to the orbifold theory on $AdS_5\times S^5/\mathbb{Z}_2$. In all cases we use the analytic bootstrap and constraints from localization to compute 1-loop and higher derivative tree level corrections to the leading supergravity approximation of the correlator. We give perturbative evidence that the localization constraint in the large $N$ and finite complexified coupling $\tau$ limit can be written for each $G$ in terms of Eisenstein series that are modular invariant in terms of $\tau_s\propto\tau$, which allows us to fix protected terms in the correlator in that limit. In all cases, we find that the flat space limit of the correlator precisely matches the type IIB S-matrix. We also find a closed form expression for the $SU(N)$ 1-loop Mellin amplitude with supergravity vertices. Finally, we compare our analytic predictions at large $N$ and finite $\tau$ to bounds from the numerical bootstrap in the large $N$ regime, and find that they are not saturated for any $G$ and any $\tau$, which suggests that no physical theory saturates these bootstrap bounds.
}
\date{}

\maketitle

\tableofcontents

\section{Introduction}
\label{intro}

Holographic correlators are correlation functions in a CFT$_d$ that are dual to scattering of particles in string theory or M-theory on AdS$_{d+1}$ times some compact manifold. Starting from \cite{Heemskerk:2009pn,Rastelli:2017udc}, they have been computed at leading order tree level\cite{Rastelli:2017ymc,Zhou:2017zaw,Caron-Huot:2018kta,Rastelli:2019gtj,Giusto:2019pxc,Goncalves:2019znr,Alday:2020lbp,Alday:2020dtb,Alday:2021odx,Zhou:2018ofp,Giusto:2020neo}, higher derivative tree level \cite{Chester:2018aca,Chester:2019pvm,Chester:2018dga,Binder:2019jwn,Binder:2019mpb,Binder:2018yvd,Binder:2021cif,Chester:2019jas,Chester:2020vyz}, and 1-loop \cite{Alday:2017xua,Aprile:2017bgs,Alday:2017vkk,Aprile:2017qoy,Aprile:2019rep,Alday:2018kkw,Alday:2019nin,Alday:2020tgi,Alday:2021ymb,Alday:2021ajh,Chester:2019pvm} in various $d$. In 4d, the simplest maximally supersymmetric example relates the stress tensor multiplet four-point function in $\mathcal{N}=4$ $SU(N)$ SYM to scattering of gravitons of type IIB string theory on $AdS_5\times S^5$. Since the complexified gauge coupling $\tau=\frac{\theta}{2\pi}+i\frac{4\pi}{g_\text{YM}^2}$ transforms under $SL(2,\mathbb{Z})$ duality invariance, the correlators can be computed at large $N$ and finite $\tau$ in terms of modular functions like Eisenstein series, which in the flat space limit precisely match the prediction from the strongly coupled string theory S-matrix \cite{Chester:2019jas,Chester:2020vyz}, which is also expected to be modular invariant in terms of the complexified string coupling $\tau_s=\chi+i/g_s$. In the small $\Im\tau$ limit, these results reproduce the large $N$ and large 't Hooft coupling $\lambda\equiv g_\text{YM}^2N$ expansion, which is dual to weakly coupled string theory, and is not sensitive to $\Re\tau=\Re\tau_s$.

In this paper, we will generalize this $\mathcal{N}=4$ SYM story to general gauge group $G=SU(N)$, $SO(2N+1)$, $SO(2N)$ or $USp(2N)$. While the $SU(N)$ case is dual to $AdS_5\times S^5$, the other three groups are dual to the orbifold theory $AdS_5\times S^5/\mathbb{Z}_2$, so from the bulk perspective there are two classes of theories: non-orbifold and orbifold. At tree level supergravity, the correlators for the two cases are identical, but they start to differ at higher derivative corrections and at loop level. As in the previously studied $SU(N)$ case, we will fix the protected $R^4$ tree level corrections using the localization constraint derived in \cite{Binder:2019jwn}, which related an integral of the correlator to derivatives $\partial_\tau\partial_{\bar\tau}\partial_m^2F(m)\vert_{m=0}$ of the mass deformed $\mathcal{N}=2^*$ free energy. This latter quantity was computed using localization in terms of a $\text{Rank}(G)$ dimensional matrix model integral \cite{Pestun:2007rz}, which we computed to any order in large $N$ and finite $\lambda\sim g_\text{YM}^2N$ following the $SU(N)$ case considered in \cite{Chester:2019pvm}. At large $\lambda$, this computes the perturbative contributions to the large $N$ finite $\tau$ limit, which we find are consistent with an expansion in Eisenstein series, as in the $SU(N)$ case. We then use the known type IIB S-matrix in the flat space limit of the holographic correlator to further constrain terms at large $N$, which confirms our answer for $R^4$ and allows us to further fix $D^4R^4$.

One novelty of the general gauge group case relative to $SU(N)$ is that the transformation under S-duality \cite{1977PhLB...72..117M,Osborn:1979tq,Witten:1978mh} is more subtle. In general, gauge groups transform to their Langlands (i.e. GNO) dual under S-duality \cite{Kapustin:2006pk}. While $SU(N)$ and $SO(2N)$ thus transform to themselves with $\tau\to-1/\tau$, up to discrete subgroups and theta angles that do not effect the quantities we consider \cite{Kapustin:2006pk,Argyres:2006qr,Aharony:2013hda},\footnote{The precise dual of $SU(N)$ is $SU(N)/\mathbb{Z}_N$, while for $SO(2N)$ it is $SO(2N)/\mathbb{Z}_2$ \cite{Kapustin:2006pk}.} $SO(2N+1)$ transforms to $USp(2N)$ with $\tau\to-1/(2\tau)$. Consistent with this non-trivial transformation, we find perturbative evidence that the $USp(2N)$ localization quantity is expanded in Eisenstein series with argument $2\tau$, as opposed to $\tau$ for the other cases. If we write these quantities in terms of $\tau_s$, however, then from the AdS/CFT dictionary $\tau_s=2\tau$ for $USp(2N)$ and $\tau_s=\tau$ for the other gauge groups, we find that results in all cases are modular invariant in terms of $\tau_s$. This had to be the case in the flat space limit where the type IIB S-matrix is modular invariant in terms of $\tau_s$, but it is remarkable that this modularity is inherited by a priori less restricted CFT quantities dual to string theory on AdS. 

We then compute the 1-loop correction coming from supergravity $R$ and $R^4$ vertices for both the orbifold and $SU(N)$ cases, using the unitarity cut method of \cite{Aharony:2016dwx}. This requires an unmixing of double trace operators formed by two half-BPS operators labeled by their superprimary dimension $p$, e.g. $p=2$ is the stress tensor. In the $SU(N)$ case there are contributions from all $p\geq2$, and our answer reproduces previous results \cite{Alday:2017xua,Aprile:2017bgs,Alday:2018pdi,Alday:2018kkw}, except that we have now been able to resum the double sum Mellin space expression of \cite{Alday:2018kkw}, see \eqref{mellin_summed}. For the orbifold case only even $p$ contributes, and the resulting 1-loop term looks very different from the $SU(N)$ case, yet in the flat space limit each case differs just by a power of 2 as expected from the orbifold factor in their bulk duals. Finally, for the 1-loop supergravity term we fix the unique contact term ambiguity using the previously described localization constraint, and find that the orbifold case differs from the $SU(N)$ case by the same factor of 2, which did not have to be the case since this contact term is subleading in the flat space limit.

One motivation for studying the orbifold theories is that their CFT data at large $N$ is bigger than the $SU(N)$ case, so if any physical theory saturates the numerical bootstrap bounds of \cite{Beem:2013qxa,Beem:2016wfs}, it would be the orbifold case for some value of $\tau$. Unfortunately, we find that our large $N$ finite $\tau$ predictions for CFT data do not saturate the bootstrap bounds at large $N$ for either case, even at the self dual point $\tau=e^{i\pi/3}$ that was conjectured to saturate the bounds in \cite{Beem:2016wfs}. Furthermore, we computed bounds using both the single correlator setup in \cite{Beem:2013qxa,Beem:2016wfs} as well as the mixed correlator setup in \cite{Bissi:2020jve} (which included $p=2,3$ half-BPS operators), and found that the upper bounds were almost indistinguishable at large $N$, even though the mixed correlator setup does not apply to the orbifold theories (since they do not contain odd $p$ half-BPS operators), so we would have expected $SU(N)$ to saturate the mixed bounds and the other cases to saturate the single correlator bounds.

Looking ahead, it would be nice to compute the other localization constraint $\partial_m^4F\vert_{m=0}$ to all orders in $1/N$ for general gauge group, as was done for $SU(N)$ in \cite{Chester:2020dja}, and to try to find the finite $N$ and $\tau$ formulae for both constraints, as has been done for $\partial_\tau\partial_{\bar\tau}\partial_m^2F(m)\vert_{m=0}$ for the $SU(N)$ case in \cite{Dorigoni:2021guq}. This will involve computing the instanton contributions to these quantities for general gauge group, which we did not consider in this paper. Finally, if we eventually want to make contact between numerical bootstrap and large $N$ analytics, it will probably be necessary to further constrain the numerical bootstrap by directly imposing the localization constraints as a function of $\tau$.

 The rest of this paper is organized as follows. In Section~\ref{4point} we review the constraints of superconformal symmetry on the stress tensor correlator, as well as its strong coupling expansion and constraints from type IIB string theory in the flat space limit.  In Section \ref{SUSYCon}, we compute the non-instanton part of $\partial_\tau\partial_{\bar\tau}\partial_m^2F(m)\vert_{m=0}$ for general gauge group both at finite $N$ and $\lambda$, and in the large $N$ and finite $g_\text{YM}$ limit (neglecting instantons), which we find is nontrivially consistent with an expansion in Eisenstein series. In Section \ref{1loop}, we compute the 1-loop amplitudes with $R$ and $R^4$ vertices, check their flat space limit, fix the 1-loop supergravity contact term ambiguity using the localization constraint, and extract CFT data. Finally, in Section \ref{numerics} we compare this large $N$ analytic data to the numerical bootstrap bounds recomputed at very high precision, and find that they do not saturate the bounds. Several technical details are given in various Appendices, and we include an attached \texttt{Mathematica} notebook with explicit results.

\section{$\mathcal{N}=4$ stress-tensor four-point function}
\label{4point}

We begin by reviewing what is already known about the stress tensor multiplet four-point function. First we discuss general constraints from the $\mathcal{N}=4$ superconformal group. Then we discuss the two different large $N$ strong coupling expansions for SYM with general gauge group in Mellin space: the expansion at finite $\tau$, as well as the more standard 't Hooft expansion at large $\lambda\sim g_\text{YM}^2 N$. Lastly, we discuss how the 10d flat space type IIB string theory S-matrix can be used to constrain the SYM correlator by taking the flat space limit.

\subsection{Basics}
\label{setup}

Let us denote the bottom component of the stress tensor multiplet by $S$.  This operator is a dimension 2 scalar in the ${\bf 20}'$ of the $SU(4)_R \cong SO(6)_R$, and can thus be represented as a rank-two traceless symmetric tensor $S_{IJ}(\vec{x})$, with indices $I, J = 1, \ldots, 6$.  For simplicity we will contract these indices with null polarization vectors $Y^I$, where $Y \cdot Y = 0$.  We are interested in studying the four-point function $\langle SSSS\rangle$, which is fixed by conformal and $SU(4)$ symmetry to take the form
 \es{2222}{
 & \langle S(\vec x_1,Y_1) S(\vec x_2,Y_2) S(\vec x_3,Y_3) S(\vec x_4,Y_4) \rangle = \frac{Y^2_{12}Y^2_{34}}{{x}_{12}^4 {x}_{34}^{4}} \mathcal{S}(U,V;\sigma,\tau)\,,
    }
where we define
 \es{uvsigmatauDefs}{
  U \equiv \frac{{x}_{12}^2 {x}_{34}^2}{{x}_{13}^2 {x}_{24}^2} \,, \qquad
   V \equiv \frac{{x}_{14}^2 {x}_{23}^2}{{x}_{13}^2 {x}_{24}^2}  \,, \qquad
   \sigma\equiv\frac{(Y_1\cdot Y_3)(Y_2\cdot Y_4)}{(Y_1\cdot Y_2)(Y_3\cdot Y_4)}\,,\qquad \tau\equiv\frac{(Y_1\cdot Y_4)(Y_2\cdot Y_3)}{(Y_1\cdot Y_2)(Y_3\cdot Y_4)} \,.
 }
 The constraints of superconformal symmetry are given by the Ward identity in \cite{Dolan:2001tt}, whose solution can be formally solved in two different ways. The first solution takes the form
  \es{T}{
 \mathcal{S}(U,V;\sigma,\tau)&=\mathcal{S}_\text{free}(U,V;\sigma,\tau)+\Theta(U,V;\sigma,\tau)\mathcal{T}(U,V)\,,\\
 \Theta(U,V;\sigma,\tau)&\equiv\tau+[1-\sigma-\tau]V+\tau[\tau-1-\sigma]U+\sigma[\sigma-1-\tau]UV+\sigma V^2+\sigma\tau U^2\,,
 }
where $\mathcal{S}_\text{free}(U,V;\sigma,\tau)$ is the free theory correlator
\es{free}{
\mathcal{S}_\text{free}(U,V;\sigma,\tau)=1+U^2\sigma^2+\frac{U^2}{V^2}\tau^2+\frac{1}{c}\left(U\sigma+\frac UV\tau+\frac{U^2}{V}\sigma\tau\right)\,,
}
so that all non-trivial interacting information is given by the $R$-symmetry invariant correlator $\mathcal{T}(U,V)$. The second solution takes the form \cite{Beem:2016wfs}
\es{redSText2}{
&\mathcal{S}(U,V;\sigma,\tau) = \Theta(U,V;\sigma,\tau)\cG(U,V)+\Phi_1(z,\bar z;\sigma,\tau)f_1(z)+\Phi_3(z,\bar z;\sigma,\tau)f_2(z)\\
&+(\Phi_1(\bar z, z;\sigma,\tau) -\Phi_2(U;\sigma,\tau))f_1(\bar z)+(\Phi_3(\bar z,z;\sigma,\tau)-\Phi_2(U;\sigma,\tau)) f_2(\bar z)+\Phi_2(U;\sigma,\tau)f_3(z)\,,\\
}
where $\Theta$ is the same as in \eqref{T} and we define
\es{Phis}{
U&=z\bar z\,,\quad V=(1-z)(1-\bar z)\,,\quad \Phi_1\equiv   \frac{z\bar z}{z-\bar z}+ \frac{z^2\bar z\tau}{(z-\bar z)( 1-z)} + \frac{z\bar z^2\sigma}{\bar z-z}\,,\quad \Phi_2\equiv  \sigma U \,,\\
\Phi_3&\equiv   \frac{z(\bar z-1)}{z-\bar z}+\frac{z^2\bar z \sigma^2 }{\bar z-z} +\frac{z^2\bar z \tau^2 + z^2\bar z(z-2)\sigma\tau + z(z+\bar z-2z\bar z)\tau}{(z-1)(z-\bar z)} +\frac{z(z+\bar z-\bar z^2)\sigma}{z-\bar z}\,.\\
}
If we then equate \eqref{redSText2} to \eqref{T}, we find that the free theory correlator fixes the holomorphic functions $f_j$ to be
\es{freef}{
f_1(z)=2+\frac1c-\frac1z+\frac{1}{z-1}\,,\qquad f_2(z)=1+\frac1c-z+\frac{1}{1-z}\,,\qquad f_3(z)=\frac1c+z+\frac1z\,, 
}
 while the remaining free part contributes to $\cG(U,V)$, which is then related to $\cT(U,V)$ as
 \es{GtoT}{
 \cT(U,V)=\cG(U,V)-\left[1+\frac{1}{V^2}+\frac1c\frac1V\right]\,.
 }
 We can expand $\cG(U,V)$ in terms of long and short multiplets as
 \es{Gexp}{
 \cG(U,V)=U^{-2}\sum_{\Delta,\ell}\lambda^2_{\Delta,\ell}G_{\Delta+4,\ell}(U,V)+\mathcal{F}^{(0)}_\text{short}(z,\bar z)+c^{-1}\mathcal{F}^{(1)}_\text{short}\,,
 }
 where $G_{\Delta,\ell}(U,V)$ with scaling dimension $\Delta$ and spin $\ell$ are 4d conformal blocks
 \es{4dblock}{
 G_{\Delta,\ell}(U,V) &=\frac{z\bar z}{z-\bar z}(k_{\Delta+\ell}(z)k_{\Delta-\ell-2}(\bar z)-k_{\Delta+\ell}(\bar z)k_{\Delta-\ell-2}( z))\,,\\
k_h(z)&\equiv z^{\frac h2}{}_2F_1(h/2,h/2,h,z)\,,
 }
 while the short multiplets OPE coefficients do not depend on the coupling and so can be computed from the free theory to give the exact expressions $\mathcal{F}_\text{short}^{(0)}$ and $\mathcal{F}_\text{short}^{(1)}$ given in \cite{Beem:2016wfs}. All non-trivial interacting information in this formulation is then given by $\Delta$ and $\ell$ for the long multiplets.

\subsection{Strong coupling expansion}
\label{strong0}

We now restrict our discussion to the case of ${\cal N} = 4$ SYM with gauge groups $SU(N)$, $SO(2N+1)$, $SO(2N)$, or $USp(2N)$. The conformal anomaly is the dimension of the group divided by four, which for each case gives
  \es{UNc}{
c_{SU(N)}=\frac{N^2-1}{4}\,,\qquad 
c_{SO(2N)}=\frac{2N^2-N}{4}\,,\qquad 
c_{SO(2N+1)}=c_{USp(2N)}=\frac{2N^2+N}{4}\,.
 }
We will consider two strong coupling limits at large $c$. In the 't Hooft limit, we define a 't Hooft coupling $\lambda$ for each gauge group as
  \es{thooft}{
\lambda_{SU(N)}&\equiv g_\text{YM}^2N\,,\qquad\qquad\qquad\, \lambda_{SO(2N)} \equiv g_\text{YM}^2\left(N-\frac14\right)\,,\\
 \lambda_{SO(2N+1)}&\equiv g_\text{YM}^2\left(N+\frac14\right)\,,\qquad \lambda_{USp(2N)} \equiv \frac12g_\text{YM}^2\left(N+\frac14\right)\,,
 }
and then consider the limit of large $c$ with fixed $\lambda$, and then large $\lambda$. We will also consider the very strong coupling limit where we keep the complexified coupling $\tau$ finite, from which we can recover the 't Hooft strong coupling limit by expanding in small $\tau$. For either strong coupling expansion, it is convenient to consider the Mellin transform $M$ of $\cT$ via
 \es{MellinDef}{
  \cT(U, V)
   = \int_{-i \infty}^{i \infty} \frac{ds\, dt}{(4 \pi i)^2} U^{\frac s2} V^{\frac t2 - 2}
    \Gamma \bigg[2 - \frac s2 \bigg]^2 \Gamma \bigg[2 - \frac t2 \bigg]^2 \Gamma \bigg[2 - \frac u2 \bigg]^2
    M(s, t) \,,
 } 
where $u \equiv 4 - s - t$.  Crossing symmetry $M(s, t) = M(t, s) = M(s, u)$ and the analytic properties of the Mellin amplitude then restrict $M(s, t)$ to have a $1/c$ and $1/\lambda$ expansion of the form (for a detailed description, see \cite{Chester:2019pvm})
\begin{align}
   {}&M =\frac1 c\Big[8M^{R}+\frac{B_0^{R^4} M^{0} }{\lambda^{\frac32}} +\frac{1}{\lambda^{\frac52}}[B_{2}^{D^4R^4} M^{2}+B_0^{D^4R^4}M^{0}] \nonumber\\
& \qquad \qquad +\frac{1}{\lambda^{3
   }}[B_{3}^{D^6R^4} M^{3}+B_2^{D^6R^4}M^{2}+B_0^{D^6R^4}M^{0}]+O(\lambda^{-\frac{7}{2}})\Big]\nonumber\\
   &+\frac{1}{c^2}
   \Big[\lambda^{\frac12} \bar{B}_0^{R^4} M^{0}+[M^{{R}|{R}}+\bar{B}^{{R}|{R}}_0M^0]+\frac{1}{\lambda}[\bar B_{3}^{D^6R^4} M^{3}+\bar B_2^{D^6R^4}M^{2}+\bar B_0^{D^6R^4}M^{0}]+O(\lambda^{-\frac32})\Big]\nonumber\\
   &+\frac{1}{c^3}
   \left[\lambda^{\frac32} \bar{\bar{B}}_2^{D^4R^4} M^{2} +\lambda\left[\bar{\bar B}_{3}^{D^6R^4} M^{3}+\bar{\bar B}_2^{D^6R^4}M^{2}+\bar{\bar B}_0^{D^6R^4}M^{0}\right]+O(\lambda^{\frac12})\right]\nonumber\\
   &+\frac{1}{c^4}
   \left[\lambda^3\left[\bar{\bar{\bar{B}}}_3^{D^6R^4} M^{3}+\bar{\bar{\bar{B}}}_2^{D^6R^4}M^{2}+\bar{\bar{\bar{B}}}_0^{D^6R^4}M^{0}\right]+O(\lambda^{\frac52})\right]+O(c^{-5})\,,
\label{MIntro}
\end{align}
where the $B$'s are numerical coefficients that cannot be fixed from symmetry alone. Here, $A|B$ refers to a term that receives contributions from a 1-loop Witten diagrams with $A,B$ vertices.\footnote{These terms can also receive contributions from other Witten diagrams that contribute at the same order, such as cubic supergravity vertices for $R|R$. Abstractly, each Mellin amplitude is simply an allowed solution to the analytic bootstrap constraints at the given order in $c$ and $\lambda$.} Terms at order $1/c^{g+1}$ correspond in the flat space limit to genus-$g$ corrections to the type IIB S-matrix in the small $g_s$ expansion.  On AdS$_5$, these terms receive contributions from $l$-loop Witten diagrams with $l\leq g$. The leading order term is tree-level supergravity \cite{Goncalves:2014ffa,Alday:2014tsa}
\es{SintM}{
M^{R}=&\frac{1}{(s-2)(t-2)(u-2)}\,,
}
whose coefficient is fixed by requiring that the unprotected R-symmetry singlet of dimension two that appears in the conformal block decomposition of the free part ${\cS}_\text{free}$ in \eqref{T} is not present in the full correlator \cite{Rastelli:2017udc}.  The $M^n$ terms in \eqref{MIntro} arise from contact Witten diagrams with vertices of the form $D^{2n}R^4$ and are degree $n$ crossing symmetric polynomials \cite{Alday:2014tsa}
 \es{treePoly}{
M^0 = 1\,\qquad M^2=s^2+t^2+u^2\,,\qquad M^{3}=stu\,.
}
The 1-loop term ${M}^{{R}|{R}}$ arises from a loop Witten diagram with two $R$ vertices and scales as $c^{-2}$, and so includes a constant contact term ambiguity. We will discuss this 1-loop term more in Section \ref{1loop}, along with other 1-loop terms ${M}^{{R}|{R^4}}$ and ${M}^{{R^4}|{R^4}}$ that scale as $c^{-2}\lambda^{-\frac52}$ and $c^{-2}\lambda^{-3}$, respectively, and so include several contact term ambiguities.  The analytic bootstrap similarly restricts the $1/c$ and finite $\tau$ expansion to be
\begin{align}
   M &=\frac8 cM^{R}+\frac{1}{c^{\frac74}} {\tilde B}_0^{R^4}(\tau,\bar\tau) M^{0} +\frac{1}{c^{2}}[M^{{R}|{R}}+\overline{B}^{{R}|{R}}_0M^0]+\frac{1}{c^{\frac94}}[{\tilde B}_{2}^{D^4R^4}(\tau,\bar\tau) M^{2}+{\tilde B}_0^{D^4R^4}(\tau,\bar\tau)M^{0}]\nonumber\\
   &\quad+\frac{1}{c^{\frac52}}[{\tilde B}_{3}^{D^6R^4}(\tau,\bar\tau) M^{3}+{\tilde B}_2^{D^6R^4}(\tau,\bar\tau)M^{2}+{\tilde B}_0^{D^6R^4}(\tau,\bar\tau)M^{0}]+O(c^{-\frac{11}{4}})\,,
\label{MIntro2}
\end{align}
where the ${\tilde B}$'s now depend on $\tau$ except the $R|R$ terms, which are the same as in the 't Hooft expansion. Next, we will discuss how some of these coefficients can be fixed by comparing to type IIB string theory in the flat space limit.

\subsection{Constraints from flat space type IIB string theory}
\label{flatCon}

The type IIB string theory S-matrix is restricted by supersymmetry to be proportional to a single function $f(s, t)$
 \es{ScattAmp}{
  {\cal A}(s, t) = \cA_{R} (s, t) f(s, t) \,,
 }
where $s,t,u=-s-t$ are 10d Mandelstam variables and $\cA_{R}$ is the tree-level four-point supergravity amplitude given in e.g. \cite{Boels:2012ie}. We can expand at small string length $\ell_s$ to get
\es{A}{
f(s,t)=1+\ell^{6}_{s}{f}_{R^4}+\ell^{8}_{s}{f}_{R|R}+\ell^{10}_{s}{f}_{D^4R^4}+\ell^{12}_{s}{f}_{D^6R^4}+\ell_s^{14}{f}_{R|R^4}+\dots\,,
}
where each coefficient depends non-trivially on the complexified string coupling $\tau_s = \chi_s + i / g_s$. The lowest few corrections are protected, and so can be computed in terms of modular functions as \cite{Green:2005ba,Green:1997as,Green:1998by,Green:1999pu}
 \es{fEisenstein}{
  f_{R^4} &= \frac{  s   t  u}{64} g_s^{\frac 32} E(\threeh, \tau_s, \bar \tau_s) \,, \\
  f_{D^4 R^4} &= \frac{ s   t   u  ( s ^2 +  t ^2 +  u ^2)}{2^{11}} g_s^{\frac 52} E(\fiveh, \tau_s, \bar \tau_s) \, ,\\
  f_{D^6 R^4} &= \frac{3(s t u)^2 }{2^{12}} g_s^{3} \cE(3,\threeh,\threeh, \tau_s, \bar \tau_s) \, .
 }
 The definition of $\cE(3,\threeh,\threeh, \tau_s, \bar \tau_s)$ is given in \cite{Green:2014yxa,Chester:2020vyz} and will not be used in this work, while the other modular functions are  Eisenstein series $  E(r, \tau_s, \bar \tau_s)$, which can be expanded at small $g_s$ as
 \es{EisensteinExpansion}{
  E(r, \tau_s, \bar \tau_s)
   ={}& \frac{2 \zeta(2 r)}{g_s^r} + 2 \sqrt{\pi} g_s^{r-1} \frac{\Gamma(r - \frac 12)}{\Gamma(r)} \zeta(2r-1) \\
    &+ \frac{2 \pi^r}{\Gamma(r) \sqrt{g_s}} \sum_{k\ne 0} \abs{k}^{r-\half}
    \sigma_{1-2r}(|k|) \, 
      K_{r - \frac 12} (2 \pi g_s^{-1} \abs{k}) \, e^{2 \pi i k \chi_s} \, ,
 }
 where the divisor sum $\sigma_p(k)$ is defined as $\sigma_p(k)=\sum_{d>0,{d|k}}  d^p$, and $K_{r - \frac 12}$ is the Bessel function of second kind. Note that each Eisenstein series includes only two perturbative in $g_s$ terms, which implies that in the small $g_s$ expansion $f_{R^4}$ and $f_{D^4R^4}$ get corrections at only genus 1 and 2, respectively. The modular function $\cE(3,\threeh,\threeh, \tau_s, \bar \tau_s)$ only has four perturbative terms \cite{Green:2014yxa,Chester:2020vyz}, so $f_{D^6R^4}$ is only corrected up to genus 3. The small $\ell_s$ expansion in 10d maps to the large $c\sim N^2$ expansion in SYM for the various gauge groups according to the dictionary
   \cite{Maldacena:1997re,Witten:1998xy,Blau:1999vz}: 
\bea
 SU(N):&\qquad \frac{L^4}{\ell^4_{s}}= {g_\text{YM}^2 N}=\lambda\,, \qquad
  &&\tau_s = \tau\,,\\
 SO(2N):&\qquad \frac{L^4}{\ell^4_{s}}=2 {g_\text{YM}^2 \Big(N-\frac14\Big)}= 2\lambda\,, \qquad
  &&\tau_s = \tau \,,\\
   SO(2N+1):&\qquad \frac{L^4}{\ell^4_{s}}=2 {g_\text{YM}^2 \Big(N+\frac14\Big)}= 2\lambda\,, \qquad
  &&\tau_s =\tau \,,\\
   USp(2N):&\qquad \frac{L^4}{\ell^4_{s}}= {g_\text{YM}^2 \Big(N+\frac14\Big)}= 2\lambda\,, \qquad
  &&\tau_s =2\tau \,.
\eea{cPlanck}
The flat space limit formula \cite{Penedones:2010ue,Fitzpatrick:2011hu,Binder:2019jwn,Chester:2020dja} then relates the Mellin amplitude to the 10d amplitude defined in \eqref{A} as
 \es{flat}{
 f(s, t) = \frac{stu}{32}\lim_{L/\ell_s \to \infty} L^{6} c \int_{\kappa-i\infty}^{\kappa+ i \infty} \frac{d\alpha}{2 \pi i} \, e^\alpha \alpha^{-6} { M} \left( \frac{L^2}{2 \alpha} s, \frac{L^2}{2 \alpha} t \right) \,,
 }
 where the dependence on the different SYM gauge groups comes from the relationship between $L$, $\ell_s$, and the CFT parameters $c$ and $\lambda$ or $\tau$ in \eqref{cPlanck}. We can apply this to $M^{R^4}$ in the finite $\tau$ expansion \eqref{MIntro2} to fix its coefficient to be
 \es{flatspaceRes}{
{\tilde B}_0^{R^4}(\tau,\bar\tau)& = \frac{15}{4 \sqrt{2 \pi^3} \orb^{\frac34}}  E(\threeh, \tau_s, \bar \tau_s)\,,
 }
where recall that the orbifold theories ($SO(2N+1)$, $SO(2N)$ and $USp(2N)$) differ from $SU(N)$ by a power of 2 due to the different AdS/CFT dictionary \eqref{cPlanck}, which we denote using the orbifold factor $\orb$, which is 1 for $SU(N)$ and 2 for the orbifold theories. Also, $\tau_s=\tau$ for all theories except $USp(2N)$ with $\tau_s=2\tau$. The 't Hooft expansion can then be recovered by looking at the perturbative terms in the Eisenstein series and writing $g_\text{YM}$ in terms of $\lambda$ and $c$ to get
 \es{flatspaceRes2}{
{ B}_0^{R^4}& = \frac{120\zeta(3)}{\orb^\frac32}\,,\qquad \bar{ B}_0^{R^4} = \frac{5}{8 \orb^\frac12}\,.
 }
The flat space constraint is not sufficient to completely fix the other Mellin amplitudes, but can be used together with other methods that we will now describe.

\section{Constraints from supersymmetric localization}
\label{SUSYCon}

There are two known constraints on the stress tensor correlator coming from supersymmetric localization. The first relates mixed $\tau$ and hypermultiplet mass $m$ derivatives of the ${\cal N} = 2^*$ sphere free energy $F=-\log Z$ to the integrated correlator in Mellin space as \cite{Binder:2019jwn,Chester:2020dja}
 \es{constraint1}{
c^2I_2[M(s,t)]=&\frac{c}{8}\frac{\partial_m^2\partial_\tau\partial_{\bar\tau}F}{\partial_\tau\partial_{\bar\tau}F}\Big\vert_{m=0}\,,\\
}
where the integral is defined as
   \es{Mints}{
I_2[M(s,t)]&\equiv-\frac18 \int_{-i \infty}^{i \infty} \frac{ds\, dt}{(2 \pi i)^2} 
\Gamma \Big[2-\frac{s}{2}\Big] \Gamma \Big[\frac{s}{2}\Big]
\Gamma \Big[2-\frac{t}{2}\Big] \Gamma \Big[\frac{t}{2}\Big]
\Gamma \Big[2-\frac{s}{2}-\frac{t}{2}\Big] \Gamma \Big[\frac{s+t}{2}\Big]M(s, t) \,.
  }
  The second constraint relates $\partial_m^4F\big\vert_{m=0}$ to a different integral of the correlator whose explicit form is given in \cite{Chester:2020dja}, but which we will not consider in detail in this paper. The $\mathcal{N}=2^*$ sphere partition function for general gauge group $G$ was computed using supersymmetric localization in \cite{Pestun:2007rz} to get
 \es{Z}{
 Z(m) =\frac{1}{|W|} \int d^r a |Z_\text{inst}(m,\tau,a)|^2e^{-\frac{8\pi^2N}{\lambda}(a,a)}\frac{1}{H(m)^r}\prod_{\alpha\in\Delta}\frac{\alpha(a)H(\alpha(a))}{H(\alpha(a)+m)}\,,
 }
 where $r$ denotes the rank of $G$, $W$ is the Weyl group, $(a,a)$ is the Killing form for the choice of basis,\footnote{The Killing form is given by $\frac{1}{2T_r}\tr_r$, where $\tr_r$ is the usual trace in the representation $r$, and $T_r$ is the Dynkin index.} $\Delta$ is the set of roots, and $H(z)$ is a product of two Barnes G-functions, namely $H(z) = e^{-(1+\gamma)z^2}G(1+ i z) G(1-i z)$.  The quantity $\abs{Z_\text{inst}(m,\tau,a)}^2$ represents the contribution to the localized partition function coming from instantons  located at the North and South poles of $S^4$ \cite{Nekrasov:2002qd,Nekrasov:2003rj,Losev:1997tp,Moore:1997dj}.  In this section, we will first compute the relevant derivatives of $Z(m)$ for $G=SU(N), SO(2N), SO(2N+1),USp(2N)$ at finite $N$ and $\lambda$ but neglecting the instanton term. Then we will expand to all orders in $1/c$ and large $\lambda$, where instantons would not contribute anyway because they are non-perturbative in the 't Hooft limit \cite{Russo:2013kea}. We will then write $\lambda$ in terms of $c$ and $\tau$ to compute the perturbative contributions to the large $c$ finite $\tau$ limit, which can be conjecturally completed to finite $\tau$ Eisenstein series for each gauge group, as was proven in the $SU(N)$ case in \cite{Chester:2019jas,Chester:2020vyz}. Finally, we will plug these results into the constraint \eqref{constraint1} to fix the $R^4$ term in the correlator, which matches the results from the flat space limit, and to fix the $D^4R^4$ term by also using the flat space limit.

\subsection{$\mathcal{N}=2^*$ sphere free energy at finite $N$}
\label{finiteN}

If we neglect the instanton term in \eqref{Z}, then we can compute $Z(m)$ in a small $m$ expansion for finite $N$ and $\lambda$ for any gauge group using the method of orthogonal polynomials \cite{mehta1981}. For instance, at $m=0$ the free energy for each gauge group is shown in Appendix \ref{locApp} to be\footnote{Since $\abs{Z_\text{inst}(0,\tau,a)}^2=1$, this result also holds in general.}
 \es{Fnom}{
 F(0)=-4c\log g_\text{YM}+\text{$g_\text{YM}$-independent}\,,
 }
 where recall that $c$ is defined differently for each gauge group. We can then simplify the quantity on the RHS of \eqref{constraint1} to be
 \es{fancyF}{
 \mathcal{F}\equiv-\frac{1}{16g_\text{YM}^4}\partial_m^2\partial_{g_\text{YM}^{-2}}^2F\big\vert_{m=0}\,,
 }
 where $\theta$ does not appear because we have ignored instantons. We can then take the mass derivatives of the localized partition function in \eqref{Z} for each gauge group to get
 \es{2m2L}{
\mathcal{F}_{SU(N)}&=\frac{g_\text{YM}^{-4}}{16}\partial_{g_\text{YM}^{-2}}^2{\sum_{i\neq j}\langle K'(a_i-a_j)\rangle}\,,\\
 \mathcal{F}_{SO(2N)}&=\frac{g_\text{YM}^{-4}}{16}\partial_{g_\text{YM}^{-2}}^2{\sum_{i\neq j} \Big[ \langle K'(a_i-a_j)\rangle+ \langle K'(a_i+a_j)\rangle \Big]}\,,\\
  \mathcal{F}_{SO(2N+1)}&=\frac{g_\text{YM}^{-4}}{16}\partial_{g_\text{YM}^{-2}}^2\Bigg[\sum_i \langle K'(a_i)\rangle+{\sum_{i\neq j} \Big[ \langle K'(a_i-a_j)\rangle+ \langle K'(a_i+a_j)\rangle \Big]}\Bigg]\,,\\
   \mathcal{F}_{USp(2N)}&=\frac{g_\text{YM}^{-4}}{16}\partial_{g_\text{YM}^{-2}}^2\Bigg[\sum_i \langle K'(2a_i)\rangle+{\sum_{i\neq j} \Big[ \langle K'(a_i-a_j)\rangle+ \langle K'(a_i+a_j)\rangle \Big]}\Bigg]\,,\\
 }
 where the expectation values are taken in terms of the zero mass partition function, and $K(z)\equiv -\frac{H'(z)}{H(z)}$ can be conveniently written in terms of its Fourier transform 
  \es{Kfourier}{
 K'(z)=-\int_0^\infty d\omega\frac{2\omega[\cos(2\omega z)-1]}{\sinh^2\omega}\,.
 }
 As shown in Appendix \ref{locApp}, the zero mass partition functions are Gaussian matrix models, so we can compute the expectation values of the exponential \eqref{Kfourier} in these ensembles using orthogonal polynomials following \cite{Fiol:2014fla} to get
\es{orthoFinal}{
\mathcal{F}_{SU(N)}&=-\int_0^\infty\frac{\omega g_\text{YM}^{-4}}{8\sinh^2\omega}\partial_{g_\text{YM}^{-2}}^2\sum_{i,j=1}^N e^{\frac{-\omega^2g_\text{YM}^2}{4\pi^2 }}\Bigg[L_{i-1}\left({\scriptstyle\frac{\omega^2g_\text{YM}^2}{4\pi^2}}\right)L_{j-1}\left({\scriptstyle\frac{\omega^2g_\text{YM}^2}{4\pi^2}}\right)\\
&\qquad\qquad\qquad\qquad\qquad\qquad\qquad-(-1)^{i-j}L_{i-1}^{j-i}\left({\scriptstyle\frac{\omega^2g_\text{YM}^2}{4\pi^2}}\right)L_{j-1}^{i-j}\left({\scriptstyle\frac{\omega^2g_\text{YM}^2}{4\pi^2}}\right)\Bigg]\,,\\
\mathcal{F}_{SO(2N)}&=-\int_0^\infty\frac{\omega g_\text{YM}^{-4}}{4\sinh^2\omega}\partial_{g_\text{YM}^{-2}}^2\sum_{i,j=1}^N e^{\frac{-\omega^2g_\text{YM}^2}{4\pi^2 }}\Bigg[L_{2(i-1)}\left({\scriptstyle\frac{\omega^2g_\text{YM}^2}{4\pi^2}}\right)L_{2(j-1)}\left({\scriptstyle\frac{\omega^2g_\text{YM}^2}{4\pi^2}}\right)\\
&\qquad\qquad\qquad\qquad\qquad\qquad\qquad-L_{2(i-1)}^{2(j-i)}\left({\scriptstyle\frac{\omega^2g_\text{YM}^2}{4\pi^2}}\right)L_{2(j-1)}^{2(i-j)}\left({\scriptstyle\frac{\omega^2g_\text{YM}^2}{4\pi^2}}\right)\Bigg]\,,\\
\mathcal{F}_{SO(2N+1)}&=-\int_0^\infty\frac{\omega g_\text{YM}^{-4}}{4\sinh^2\omega}\partial_{g_\text{YM}^{-2}}^2\Bigg( e^{\frac{-\omega^2g_\text{YM}^2}{4\pi^2 }}\sum_{i,j=1}^N\Bigg[L_{2i-1}\left({\scriptstyle\frac{\omega^2g_\text{YM}^2}{4\pi^2}}\right)L_{2j-1}\left({\scriptstyle\frac{\omega^2g_\text{YM}^2}{4\pi^2}}\right)\\
&\qquad-L_{2i-1}^{2(j-i)}\left({\scriptstyle\frac{\omega^2g_\text{YM}^2}{4\pi^2}}\right)L_{2j-1}^{2(i-j)}\left({\scriptstyle\frac{\omega^2g_\text{YM}^2}{4\pi^2}}\right)\Bigg]+e^{\frac{-\omega^2g_\text{YM}^2}{8\pi^2 }}\sum_{i=1}^N L_{2i-1}\left({\scriptstyle\frac{\omega^2g_\text{YM}^2}{4\pi^2}}\right)\Bigg)\,,\\
\mathcal{F}_{USp(2N)}&=-\int_0^\infty\frac{\omega g_\text{YM}^{-4}}{4\sinh^2\omega}\partial_{g_\text{YM}^{-2}}^2\Bigg( e^{\frac{-\omega^2g_\text{YM}^2}{8\pi^2 }}\sum_{i,j=1}^N\Bigg[L_{2i-1}\left({\scriptstyle\frac{\omega^2g_\text{YM}^2}{8\pi^2}}\right)L_{2j-1}\left({\scriptstyle\frac{\omega^2g_\text{YM}^2}{8\pi^2}}\right)\\
&\qquad-L_{2i-1}^{2(j-i)}\left({\scriptstyle\frac{\omega^2g_\text{YM}^2}{8\pi^2}}\right)L_{2j-1}^{2(i-j)}\left({\scriptstyle\frac{\omega^2g_\text{YM}^2}{8\pi^2}}\right)\Bigg]+e^{\frac{-\omega^2g_\text{YM}^2}{4\pi^2 }}\sum_{i=1}^N L_{2i-1}\left({\scriptstyle\frac{\omega^2g_\text{YM}^2}{2\pi^2}}\right)\Bigg)\,,\\
}
where $L_i^j(z)$ are generalized Laguerre polynomials, and the $SU(N)$ case was already computed in \cite{Chester:2019pvm}. For $SO(2N+1)$, the result holds for $N>1$, while for $SO(3)$ we must rescale the quantity in parentheses by $g_\text{YM}\to\sqrt{2}g_\text{YM}$, where the difference comes from the discontinuity of the standard Killing form $(a,a)$, i.e. of the Dynkin index, for $SO(N)$ with $N=3$ and $N>3$. One can check that these formulae satisfy the Lie algebra isomorphisms $SU(2)\cong SO(3)\cong USp(2)$, $SU(4)\cong SO(6)$, $SU(2)\times SU(2)\cong SO(4)$, and $SO(5)\cong USp(4)$.

\subsection{$1/N$ expansions}
\label{largeN}

We can now take the large $N$ limit of these finite $N$ results. From the $SU(N)$ case considered in \cite{Chester:2019pvm}, which was computed to all orders in $1/N$ using topological recursion for $U(N)$ Gaussian matrix models \cite{Eynard:2004mh,Eynard:2008we}, we expect that the finite $\lambda$ answer should be written in terms of Bessel functions. Since the topological recursion relations have not yet been derived for the other gauge groups, we will instead directly expand the finite $N$ and $\lambda$ result in \eqref{orthoFinal}. We do this by first considering the weak coupling expansion, where we define the weak coupling limit 't Hooft coupling for each gauge group as
\es{weakLam}{
\lambda^{SU(N)}_w&=Ng_\text{YM}^2\,,\qquad\qquad\quad\; \,\lambda^{SO(2N)}_w=2Ng_\text{YM}^2\,,\\
 \lambda^{SO(2N+1)}_w&=(2N+1)g_\text{YM}^2\,,\qquad  \lambda^{USp(2N)}_w=\frac12(2N+1)g_\text{YM}^2\,.
}
We then perform the sums over $i,j$ in \eqref{orthoFinal} at each order in $\lambda_w$, take the large $N$ expansion, and finally resum $\lambda_{w}$ at each order in $1/N$. To order $O(N^0)$ we found
 \es{UNanswer}{
&\cF_{SU(N)}=N^2\int_{0}^\infty d\omega\, \omega \frac{J_1(\frac{\sqrt{\lambda_w}}{\pi}\omega)^2-J_2(\frac{\sqrt{\lambda_w}}{\pi}\omega)^2}{4\sinh^2\omega}+O(N^{0})\,,\\
}
and
  \es{SO2Nanswer}{
& \mathcal{F}_{SO(2N)}=\int_0^\infty \frac{d\omega}{\sinh^2\omega} \Bigg[\frac{N^2}{2}  \Big(\frac{\left(\lambda_w  \omega^2-4 \pi ^2\right)
   J_1\left({\scriptstyle\frac{\omega \sqrt{\lambda_w }}{\pi }}\right)^2}{\lambda_w  \omega}-\omega
   J_0\left({\scriptstyle\frac{\omega \sqrt{\lambda_w }}{\pi }}\right)^2+\frac{4 \pi 
   J_1\left({\scriptstyle\frac{\omega \sqrt{\lambda_w }}{\pi }}\right) J_0\left({\scriptstyle\frac{\omega \sqrt{\lambda_w }}{\pi }}\right)}{\sqrt{\lambda_w }}\Big)\\
   &+\frac{ N\omega}{16 \pi } \left(2
   \pi  J_1\left({\scriptstyle\frac{\omega \sqrt{\lambda_w }}{\pi }}\right)^2+\sqrt{\lambda_w } w
   \left(1-4 J_0\left({\scriptstyle\frac{\omega \sqrt{\lambda_w }}{\pi }}\right)\right)
   J_1\left({\scriptstyle\frac{\omega \sqrt{\lambda_w }}{\pi }}\right)-2 \sqrt{\lambda_w } \omega
   J_1\left({\scriptstyle \frac{2 \omega \sqrt{\lambda_w }}{\pi }}\right)\right)\Bigg]+O(N^0)\,,\\
    }
 and
    \es{SO2N3answer}{
 & \mathcal{F}_{SO(2N+1)}=\int_0^\infty \frac{d\omega}{\sinh^2\omega} \Bigg[  \frac{N^2}{2}  \Big(\frac{\left(\lambda_w \omega^2-4 \pi ^2\right)
   J_1\left({\scriptstyle\frac{\omega \sqrt{\lambda_w }}{\pi }}\right)^2}{\lambda_w \omega}-\omega J_0\left({\scriptstyle\frac{\omega \sqrt{\lambda_w }}{\pi }}\right)^2+\frac{4 \pi  J_1\left({\scriptstyle\frac{\omega \sqrt{\lambda_w }}{\pi }}\right) J_0\left({\scriptstyle\frac{\omega \sqrt{\lambda_w }}{\pi }}\right)}{\sqrt{\lambda_w}}\Big)\\
   &-\frac{N}{16 \pi \lambda_w \omega}\Big(2 \lambda_w^{3/2} \omega^3
   J_1\left({\scriptstyle\frac{2\omega \sqrt{\lambda_w }}{\pi }}\right)+8 \pi 
   \lambda_w \omega^2 J_0\left({\scriptstyle\frac{\omega \sqrt{\lambda_w}}{\pi}}\right)^2-{{  \lambda^{\frac32}_w   }  \omega ^3 J_1\left({\scriptstyle\frac{\omega \sqrt{ \lambda_w  }   }{\pi  }}\right)}\\
   &+4 \sqrt{\lambda_w} \omega \left(\lambda_w \omega^2-8
   \pi ^2\right) J_0\left({\scriptstyle\frac{\omega \sqrt{\lambda_w }}{\pi }}\right)
   J_1\left({\scriptstyle\frac{\omega \sqrt{\lambda_w }}{\pi }}\right)+2 \left(16 \pi
   ^3-5 \pi  \lambda_w \omega^2\right) J_1\left({\scriptstyle\frac{\omega\sqrt{\lambda_w}}{\pi }}\right)^2\Big)  \Bigg]+O(N^0)\,,
}
and
 \es{USPNanswer}{
 & \mathcal{F}_{USp(2N)}=\int_0^\infty \frac{d\omega}{\sinh^2\omega} \Bigg[  \frac{N^2}{2}  \Big(\frac{\left(\lambda_w \omega^2-4 \pi ^2\right)
   J_1\left({\scriptstyle\frac{\omega \sqrt{\lambda_w }}{\pi }}\right)^2}{\lambda_w \omega}-\omega J_0\left({\scriptstyle\frac{\omega \sqrt{\lambda_w }}{\pi }}\right)^2+\frac{4 \pi  J_1\left({\scriptstyle\frac{\omega \sqrt{\lambda_w }}{\pi }}\right) J_0\left({\scriptstyle\frac{\omega \sqrt{\lambda_w }}{\pi }}\right)}{\sqrt{\lambda_w}}\Big)\\
   &-\frac{N}{16 \pi \lambda_w \omega}\Big(\lambda_w^{3/2} \omega^3 J_1\left({\scriptstyle\frac{\omega \sqrt{\lambda_w }}{\pi }}\right)+2 \lambda_w^{3/2} \omega^3
   J_1\left({\scriptstyle\frac{2\omega \sqrt{\lambda_w }}{\pi }}\right)+8 \pi 
   \lambda_w \omega^2 J_0\left({\scriptstyle\frac{\omega \sqrt{\lambda_w}}{\pi}}\right)^2-4{{  \lambda^{\frac32}_w   }  \omega ^3 J_1\left({\scriptstyle\frac{2\omega \sqrt{ \lambda_w  }   }{\pi  }}\right)}\\
   &+4 \sqrt{\lambda_w} \omega \left(\lambda_w \omega^2-8
   \pi ^2\right) J_0\left({\scriptstyle\frac{\omega \sqrt{\lambda_w }}{\pi }}\right)
   J_1\left({\scriptstyle\frac{\omega \sqrt{\lambda_w }}{\pi }}\right)+2 \left(16 \pi
   ^3-5 \pi  \lambda_w \omega^2\right) J_1\left({\scriptstyle\frac{\omega\sqrt{\lambda_w}}{\pi }}\right)^2\Big)  \Bigg]+O(N^0)\,,
}
where $\lambda_w$ is defined in each case in \eqref{weakLam}. Note that the $SU(N)$ case only includes even powers of $N$ and pairs of Bessel functions, while the other cases include all powers of $N$ and single Bessel functions. This pattern continues at subleading orders in $1/N$, which are given in the attached \texttt{Mathematica} file. For $SU(N)$, these results match those computed from topological recursion in \cite{Chester:2019pvm}.

 The next step is to compute the strong coupling expansion by looking at large 't Hooft coupling. For this we want to use the 't Hooft coupling that is naturally related to the AdS radius, which is the $\lambda$ defined for each gauge group in \eqref{thooft}.\footnote{For $SU(N)$, the strong and weak coupling definitions coincide, but for the other cases they differ.} To do the large $\lambda$ expansion, we use the Mellin-Barnes formulae for Bessel functions:
\es{mbbessel}{
J_\mu(x)J_\nu(x) &= \frac 1 {2\pi i}\int_{c-\infty i}^{c+\infty i} ds\frac{\Gamma(-s)\Gamma(2s+\mu+\nu+1)\left(\frac12 x\right)^{\mu+\nu+2s}}{\Gamma(s+\mu+1)\Gamma(s+\nu+1)\Gamma(s+\mu+\nu+1)} \,,\\
J_\mu(x) &= \frac 1 {2\pi i}\int_{c-\infty i}^{c+\infty i}ds\frac{\Gamma(-s)
\left(\frac12 x\right)^{\mu+2s}}{\Gamma(s+\mu+1)} \,,\\
}
perform the $\omega$ integrals using
\es{wint}{
\int_0^\infty d\omega\ \frac{\omega^{a}}{\sinh^2\omega} = \frac{1}{2^{a-1}}\Gamma(a+1)\zeta(a) \,,
}
 and finally close the $s$ contours to the left (closing to the right would give the weak coupling expansion). For $SU(N)$ we get
 \es{tohooftSUN}{
 \mathcal{F}_{SU(N)}&= c \left(\frac14-\frac{3 \zeta (3)}{\lambda ^{3/2}}+\frac{45 \zeta
   (5)}{4 \lambda ^{5/2}}+O(\lambda^{-\frac72})\right)+\Big(-\frac{\sqrt{\lambda
   }}{64 }+\frac{1}{16}+O(\lambda^{-\frac32})\Big)\\
 & \qquad +\frac1c\Big(\frac{{\lambda^{\frac32}
   }}{24576 }+O(\lambda^{\frac12})\Big)+O(c^{-2})\,,\\
   }
   where we wrote $N$ in terms of $c$ using \eqref{UNc}, which is where the difference between $U(N)$ and $SU(N)$ finally enters, and these results recover those in \cite{Chester:2019pvm}. For $SO(2N)$, $SO(2N+1)$ and $USp(2N)$, after using the different definitions of $c$ in \eqref{UNc} and $\lambda$ in \eqref{thooft} for each case, we find essentially the same answer 
    \es{tohooftSON}{
\mathcal{F}_{orb}&= c \left(\frac14-\frac{3 \zeta (3)}{2 \sqrt{2} \lambda ^{3/2}}+\frac{45 \zeta
   (5)}{16 \sqrt{2} \lambda ^{5/2}}+O(\lambda^{-\frac72})\right)+\Big(-\frac{\sqrt{\lambda
   }}{64 \sqrt{2}}+\frac{1}{128}+O(\lambda^{-\frac32})\Big)\\
   & \qquad +\frac1c\Big(\frac{{\lambda^{\frac32}
   }}{24576\sqrt{2} }+O(\lambda^{\frac12})\Big)+O(c^{-2})\,,
 }
 which involved nontrivial cancellations so that other powers of $c$ and $\lambda$ would not appear. For all gauge groups we only find half integer powers of $\lambda$, unlike $\partial_m^4F\big\vert_{m=0}$ as computed for $SU(N)$ in \cite{Chester:2020dja}, which included both integer and half integer powers of $\lambda$.
  
  Finally, we consider the large $c$ and finite $\tau$ limit. We can compute the terms which are perturbative in $g_\text{YM}$ from the 't Hooft limit by simply expressing $\lambda$ in terms of $c$ and $g_\text{YM}$ using the different definitions for each gauge group, and then reexpanding at large $c$. By expanding \eqref{tohooftSUN} and \eqref{tohooftSON} to several more orders in this manner using the expressions in the attached \texttt{Mathematica} notebook, we found that they are nontrivially consistent with the Eisenstein series expansions
   \es{finiteNSUN}{
& \mathcal{F}_{SU(N)}=\frac{c}{4}+\frac{1}{16}-c^{\frac14}\frac{3
   E(\threeh,\tau,\bar\tau)}{32 \sqrt{2} \pi^{\frac32}}+\frac{1}{c^{\frac14}}\frac{45
  E(\fiveh,\tau,\bar\tau) }{1024 \sqrt{2}  \pi^{\frac52}}\\
 & \; +\frac{1}{c^{\frac34}} \Bigg[\frac{4725E(\sevenh,\tau,\bar\tau)}{262144 \sqrt{2}  \pi^{\frac72}}-\frac{423
  E(\threeh,\tau,\bar\tau)  }{65536 \sqrt{2}  \pi^{\frac{3}{2}}}\Bigg]+\frac{1}{c^{\frac54}} \Bigg[\frac{99225E(\nineh,\tau,\bar\tau)}{4194304 \sqrt{2}  \pi^{\frac92}}-\frac{4005
  E(\fiveh,\tau,\bar\tau)}{1048576 \sqrt{2}  \pi^{\frac52}}\Bigg]\\
 &\;+\frac{1}{c^{\frac74}}  \Bigg[\frac{245581875E(\elevenh,\tau,\bar\tau)}{4294967296 \sqrt{2} \pi
   ^{\frac{11}{2}}} -\frac{6440175E(\sevenh,\tau,\bar\tau)}{1073741824 \sqrt{2}  \pi^{\frac72}}+ \frac{93303E(\threeh,\tau,\bar\tau)}{134217728 \sqrt{2}  \pi^{\frac{3}{2}}} \Bigg]+O(c^{-\frac94})\,,\\
   }
   and
\begin{align}
&\mathcal{F}_{orb}=\frac{c}{4}+\frac{1}{128}-c^{\frac14}\frac{3  E(\threeh,\tau_s,\bar\tau_s)}{64\cdot
  {2}^{\frac14}  \pi^{\frac{3}{2}}}+\frac{1}{c^{\frac14}}\frac{45 E(\fiveh,\tau_s,\bar\tau_s)  }{2048\cdot
   2^{\frac34}  \pi^{\frac52}}
\label{finiteNSON}\\
   &+\frac{1}{c^{\frac34}}  \Bigg[  \frac{4725 E(\sevenh,\tau_s,\bar\tau_s)   }{1048576\cdot 2^{\frac14}  \pi^{\frac72}}-\frac{207
   E(\threeh,\tau_s,\bar\tau_s)}{262144\cdot 2^{\frac14}  \pi^{\frac{3}{2}}}   \Bigg]
  +\frac{1}{c^{\frac{5}{4}}} \Bigg[  \frac{99225 E(\nineh,\tau_s,\bar\tau_s)}{16777216\cdot 2^{\frac34}  \pi^{\frac92}}-\frac{4545
   E(\fiveh,\tau_s,\bar\tau_s)}{4194304\cdot 2^{\frac34}  \pi^{\frac52}}   \Bigg]
\nonumber\\
&+\frac{1}{c^{\frac74}} \Bigg[ \frac{245581875 E(\elevenh,\tau_s,\bar\tau_s)}{34359738368\cdot 2^{\frac14}
    \pi^{\frac{11}{2}}} -\frac{11656575 E(\sevenh,\tau_s,\bar\tau_s)}{8589934592\cdot 2^{\frac14}  \pi^{\frac72}} + \frac{55503 E(\threeh,\tau_s,\bar\tau_s)}{1073741824\cdot 2^{\frac14}  \pi^{\frac{3}{2}}} \Bigg]  +O(c^{-\frac94})\,,
\nonumber
\end{align}
 where we used the perturbative terms in the Eisenstein expansion \eqref{EisensteinExpansion}, and recall that $\tau_s=\tau$ for all gauge groups except $USp(2N)$ where $\tau_s=2\tau$.  Note that there are only two perturbative terms in the Eisenstein series expansion, which corresponds to the fact that there is a finite number of perturbative terms at each order in $1/c$. For $SU(N)$, this recovers the result that was rigorously derived in \cite{Chester:2019jas,Chester:2020vyz}.

\subsection{Constraining the stress-tensor correlator}
\label{locCon}

We can now input the localization results for $\partial_m^2\partial_\tau\partial_{\bar\tau}F\big\vert_{m=0}$ into the integrated constraint \eqref{constraint1} to constrain the stress tensor correlator as expanded at large $N$ in \eqref{MIntro} and \eqref{MIntro2}. For the degree $n$ polynomial Mellin amplitudes $M^n$ in \eqref{treePoly} as well as the tree level supergravity amplitude $M^R$ in \eqref{SintM}, we can compute the integrals in \eqref{Mints} using Barnes' Lemma
\es{barnes}{
\int_{-i\infty}^{i\infty}\frac{ds}{2\pi i}\Gamma(a+s)\Gamma(b+s)\Gamma(c-s)\Gamma(d-s) = \frac{\Gamma(a+c)\Gamma(b+d)\Gamma(b+c)\Gamma(b+d)}{\Gamma(a+b+c+d)} \,,
}
to get \cite{Binder:2019jwn,Chester:2020dja}
\es{integrals}{
I_2[M^{R}]&=\frac{1}{32}\,,\quad\qquad I_2[M^0]=-\frac{1}{40}\,,\qquad I_2[M^2]=-\frac{2}{35}\,,\qquad I_2[M^3]=-\frac{4}{35}\,.\\
}
The localization constraint \eqref{constraint1} then fixes the $R^4$ coefficients to the values in \eqref{flatspaceRes} and \eqref{flatspaceRes2} that were independently fixed from the flat space limit, which is a check of AdS/CFT at genus-1 for general gauge group, generalizing the $SU(N)$ case first shown in \cite{Binder:2019jwn}. We can then combine this localization constraint with the flat space limit constraint to rigorously fix the $D^4R^4$ coefficients in the 't Hooft limit to be
 \es{locRes2}{
{ B}_2^{D^4R^4}& = -\frac13{ B}_0^{D^4R^4}= \frac{630\zeta(5)}{\orb^{\frac52}}\,,\qquad \bar{ \bar B}_2^{D^4R^4} =-\frac13 \bar{ \bar B}_0^{D^4R^4} =- \frac{7}{3072 \orb^\frac12}\,.
 }
We can also use the conjectured localization expressions in \eqref{finiteNSUN} and \eqref{finiteNSON} for the large $c$ and finite $\tau$ limit to get
 \es{locRes}{
{\tilde B}_2^{D^4R^4}(\tau_s,\bar\tau_s)&= -\frac13{\tilde B}_0^{D^4R^4}(\tau_s,\bar\tau_s) = \frac{315}{128 \sqrt{2 \pi^5} \orb^\frac54}  E(\fiveh, \tau_s, \bar \tau_s)\,.
 }
 For both the $R^4$ and $D^4R^4$ cases, we observe that the finite $\tau$ localization results for $SU(N)$ \eqref{finiteNSUN} and the orbifold cases \eqref{finiteNSON} differ by a factor of $2^{\frac{l}{8}}$, where $l$ is the $\ell_s$ scaling of the corresponding term in the type IIB S-matrix \eqref{A}. This is the same factor that appeared in the constraint from the flat space limit formula \eqref{flat}, which arises from the different AdS/CFT dictionary \eqref{cPlanck} for each case, and is ultimately explained by the orbifold factor in the bulk duals. This suggests that for all tree level terms\footnote{Since $c$ is the only expansion parameter at finite $\tau$, from the bulk perspective there is in general no difference between tree and loop level. At low orders, however, one can distinguish between tree and loop terms by the different power of $c$ that multiply them.}, the difference between the orbifold and non-orbifold cases will simply be this orbifold factor, as was similarly observed for M-theory duals in \cite{Alday:2020tgi,Alday:2021ymb}. For $SU(N)$, the $D^6R^4$ coefficients in \eqref{flatspaceRes} and \eqref{flatspaceRes2} were fixed in \cite{Chester:2020dja,Chester:2020vyz} using the localization constraint from $\partial_m^4F\big\vert_{m=0}$ combined with the $\partial_m^2\partial_\tau\partial_{\bar\tau}F\big\vert_{m=0}$ constraint and the flat space limit. We can then divide by $2^{\frac{l}{8}}$ for $l=12$ to conjecturally obtain the orbifold results from the known $SU(N)$ results as
  \es{locResConj}{
{\tilde B}_3^{D^6R^4}(\tau_s,\bar\tau_s)&= -4{\tilde B}_2^{D^6R^4}(\tau_s,\bar\tau_s) = -\frac14{\tilde B}_0^{D^6R^4}(\tau_s,\bar\tau_s) = \frac{945\cE(3,\threeh,\threeh, \tau_s, \bar \tau_s)}{64\pi^3 \orb^\frac32}\,.
 }
The 't Hooft limit coefficients in \eqref{flatspaceRes} can then be extracted from the perturbative part of the modular function as given in \cite{Chester:2020dja,Chester:2020vyz} to get
  \es{locResConj2}{
{{{ B}}}_3^{D^6R^4}& = -4{{{ B}}}_2^{D^6R^4}=-\frac14{{{ B}}}_0^{D^6R^4}= \frac{5040\zeta(3)^2}{\orb^3}\,,\\
 {{{\bar B}}}_3^{D^6R^4}& = -4{{{\bar B}}}_2^{D^6R^4}=-\frac14{{{\bar B}}}_0^{D^6R^4}= \frac{105\zeta(3)}{4 \orb^2}\,,\\
 {\bar{{\bar B}}}_3^{D^6R^4}& = -4{\bar{{\bar B}}}_2^{D^6R^4}=-\frac14{\bar{{\bar B}}}_0^{D^6R^4}=\frac{21}{256 \orb}\,,\\
  \bar{\bar{{\bar B}}}_3^{D^6R^4}& = -4\bar{\bar{{\bar B}}}_2^{D^6R^4}=-\frac14\bar{\bar{{\bar B}}}_0^{D^6R^4}= \frac{1}{221184}\,.
 }
 We will explore more constraints from localization and the flat space limit in the next section, where we consider 1-loop terms.

\section{Stress-tensor correlator at 1-loop }
\label{1loop}

We will now compute the 1-loop terms with $R$ and $R^4$ vertices in the large $c$ expansion for the $SU(N)$ and other gauge groups, dual to $AdS_5\times S^5$ and $AdS_5\times S^5/\mathbb{Z}_2$, respectively. Since the results for the 't Hooft expansion can be trivially obtained from the finite $\tau$ expansion, as discussed in previous sections, we will only show the finite $\tau$ expansion in this section. In particular the $R|R$, $R|R^4$, and $R^4|R^4$ terms appear at orders $c^{-2}$, $c^{-\frac{11}{4}}$, and $c^{-\frac{7}{2}}$, respectively. We obtain the 1-loop terms by first computing the 1-loop double-discontinuity (DD) from tree and GFFT data, and then using it to write the entire correlator in Mellin space using crossing symmetry up to contact term ambiguities. We then compare the correlators for both theories to the relevant 1-loop corrections to the 10d S-matrix in the flat space limit, and find a precise match for 1-loop amplitudes on both $AdS_5\times S^5$ and $AdS_5\times S^5/\mathbb{Z}_2$. For $R|R$, we use the localization constraints from the previous section to fix the unique contact term ambiguity for each theory, and find that they are simply related by the orbifold factor. Finally, we extract low-lying CFT data using two methods: the Lorentzian inversion integral applied to the DD that does not apply to certain low values of spin \cite{Alday:2016njk,Caron-Huot:2017vep}, and a projection method applied to the entire Mellin amplitude \cite{Heemskerk:2009pn,Chester:2018lbz}. Each method agrees in general, while only the projection method can be used to extract the low spin data that is affected by the contact term ambiguities, which in some cases we fixed using localization.

\subsection{One-loop from tree level}
\label{1loopfrom}

Since the short multiplets are all $1/c$ exact, only long multiplets will appear at 1-loop. These long multiplets are double trace operators that can be written as $S_p\partial_{\mu_1}\dots\partial_{\mu_\ell}(\partial^2)^nS_p$, where $S_p$ is the $\Delta=p$ bottom component of the single trace half-BPS multiplet dual to the $p$th lowest KK mode, so that the stress tensor is $S_2$. The long multiplets thus have spin $\ell$ and twist $ t\equiv \Delta-\ell=2p+2n$ for integer $n$ so for $t\geq2$ there are $t/2$ such degenerate operators due to the different ways of adding $p$ and $n$ to get the same twist, which we label using the degeneracy label $I$. We can expand the CFT data of these long operators as 
\es{anomA}{
\Delta_{t,\ell,I}&=t+\ell+c^{-1}\gamma^{R}_{t,\ell,I}+c^{-\frac74}\gamma^{R^4}_{t,\ell,I}+c^{-2}\gamma^{R|R}_{t,\ell,I}+\dots\,,\\
(\lambda_{p,t,\ell,I})^2&=(\lambda^{(0)}_{p,t,\ell,I})^2+c^{-1}(\lambda^{R}_{p,t,\ell,I})^2+c^{-\frac74}(\lambda^{R^4}_{p,t,\ell,I})^2+c^{-2}(\lambda^{R|R}_{p,t,\ell,I})^2+\dots\,,\\
}
where $\lambda_{p,t,\ell,I}$ is the OPE coefficient in $S_p\times S_p$. At GFFT and tree level, the only difference between the $AdS_5\times S^5$ and $AdS_5\times S^5/\mathbb{Z}_2$ theories is that the orbifold projects out all long multiplets constructed from $S_p$ with odd $p$, while at loop level the values of the CFT data themselves will differ. We can then apply this large $c$ expansion to the block expansion in \eqref{Gexp} of the position space correlator $\cG(U,V)$ to get $R|R$ at order $c^{-2}$:
\es{RR}{
\cG^{R|R}=&\sum_{t=4,6,\dots}\sum_{\ell\in\text{Even}}\Big[\frac18\langle(\lambda^{(0)}_{t,\ell})^2(\gamma^R_{t,\ell})^2\rangle(\log^2U+4\log U\partial_t^\text{no-log}+4(\partial_t^\text{no-log})^2) \\
&+\frac12\langle(\lambda^{R})^2_{t,\ell}\gamma^{R}_{t,\ell}\rangle(\log U+2\partial_t^\text{no-log})\\
 &+\frac12\langle(\lambda^{(0)}_{t,\ell})^2\gamma^{{R}|{R}}_{t,\ell}\rangle(\log U+2\partial_t^\text{no-log})+\langle(\lambda^{{R}|{R}}_{t,\ell})^2\rangle\Big] U^{-2}G_{t+\ell+4,\ell}(U,V)\,,
}
where we suppressed the $p=2$ subscripts for simplicity,  $\langle\rangle$ denotes the sum over the degeneracy label $I$ in \eqref{anomA}, and $\partial_t^\text{no-log}G_{t+\ell+4,\ell}$ means that after taking the derivative we consider the term that does not include a $\log U$, since the terms with a log have already been written separately. The expression for $\cG^{R^4|R^4}$ at order $c^{-\frac{7}{2}}$ is identical except we replace $R\to R^4$ and the sum for the long multiplets is now restricted to $\ell=0$, since tree $R^4$ only contributes to $\ell=0$ CFT data. The expression for $\cG^{R|R^4}$ at order $c^{-\frac{11}{4}}$ is similar except we replace the $\frac18$ in the first line by $\frac14$ since the vertices are different.

As shown in \cite{Aharony:2016dwx}, the entire 1-loop term up to the contact term ambiguities described in Section \ref{strong0} can in fact be constructed from the $\log^2 U$ terms shown above, which are written in terms of GFFT and tree data, since under $1\leftrightarrow3$ crossing
\es{crossing}{
 V^2 \cG(U,V)-U^2\cG(V,U)+(U-V)\frac1c+U^2-V^2=0\,,
}
these are related to $\log^2V$ terms that are the only contributions at this order to the DD, which can be used to reconstruct the full 1-loop correlator as shown in \cite{Alday:2016njk,Caron-Huot:2017vep}. A subtlety is that the $I$ sum $\langle(\lambda^{(0)}_{t,\ell})^2\gamma^A_{t,\ell}\gamma^{B}_{t,\ell}\rangle$ for 1-loop vertices $A,B$ is what appears in the $\log^2U$ term, whereas the different sums $\langle(\lambda^{(0)}_{t,\ell})^2\gamma^A_{t,\ell}\rangle$ and $\langle(\lambda^{(0)}_{t,\ell})^2\gamma^B_{t,\ell}\rangle$ are what appear at tree level. As shown in Appendix A of \cite{Alday:2018pdi}, one can compute $\langle(\lambda^{(0)}_{t,\ell})^2\gamma^A_{t,\ell}\gamma^{B}_{t,\ell}\rangle$ from GFFT $\langle S_pS_pS_qS_q\rangle$ and tree level $\langle S_2S_2S_pS_p\rangle$ data as
\es{appA}{
\langle(\lambda^{(0)}_{t,\ell})^2\gamma^A_{t,\ell}\gamma^{B}_{t,\ell}\rangle=\sum_{p=2}^{t/2}\frac{\langle\lambda^{(0)}_{2,t,\ell}\lambda^{(0)}_{p,t,\ell}\gamma^A_{t,\ell}\rangle  \langle\lambda^{(0)}_{2,t,\ell}\lambda^{(0)}_{p,t,\ell}\gamma^B_{t,\ell}\rangle}{ {\langle(\lambda^{(0)}_{p,t,\ell})^2\rangle} }\,,
}
where we summed over each $p$ where a given twist $t$ long multiplet appears. For $SU(N)$, this includes all integer $p$, while for the orbifold cases it only includes even $p$. The GFFT average OPE coefficients were extracted from $\langle S_pS_pS_pS_p\rangle$ in e.g. \cite{Aprile:2017xsp}\footnote{We include a factor of $1/p^2$ that was missing from \cite{Aprile:2017xsp}.}
\es{ppppLam}{
&\langle (\lambda^{(0)}_{p,t,\ell})^2\rangle\equiv \sum_{I=1}^{t/2}( \lambda^{(0)}_{p,t,\ell,I})^2=\frac{24  (\ell+1) \left(\frac{t }{2}-2\right)! \left(\frac{t }{2}!\right)^2  (\ell+t +2)
   \left(\ell+\frac{t }{2}-1\right)!}
   {\left(\frac{t }{2}+2\right)!
   t ! p^2 (p+1) (p-2)! ((p-1)!)^3 }\\
&\qquad\qquad\qquad\qquad\qquad\qquad\times\frac{ \left(\left(\ell+\frac{t }{2}+1\right)!\right)^2 \left(p+\frac{t
   }{2}\right)! \left(\ell+p+\frac{t }{2}+1\right)!}{ \left(\ell+\frac{t }{2}+3\right)! (2 \ell+t +2)! \left(\frac{t }{2}-p\right)! \left(\ell-p+\frac{t
   }{2}+1\right)!}\,,
}
while the analogous $\langle \lambda^{(0)}_{p,t,\ell}\lambda^{(0)}_{q,t,\ell}\rangle$ vanishes for $p\neq q$ since $\langle S_pS_pS_qS_q\rangle$ is trivial in GFFT. The supergravity average anomalous dimensions were extracted from $\langle S_2S_2S_pS_p\rangle$ in \cite{Aprile:2017xsp} to get in our conventions
\es{averageSG}{
\langle\lambda^{(0)}_{2,t,\ell}\lambda^{(0)}_{p,t,\ell}\gamma^R_{t,\ell}\rangle=
\frac{\pi  (-1)^{p+1}  \Gamma \left(\ell+\frac{t }{2}+2\right) \left(-p+\frac{t
   }{2}+1\right)_p \Gamma \left(p+\frac{t }{2}+1\right)
}{
2^{2 \ell+2 t +1} \Gamma (p-1) \Gamma (p) \Gamma \left(\frac{t
   +1}{2}\right) \Gamma \left(\ell+\frac{t }{2}+\frac{3}{2}\right)}\,,
}
which is the same for orbifold and non-orbifold. The $R^4$ expression depends on the theory, and was extracted for the $SU(N)$ theory in the 't Hooft limit in \cite{Alday:2018pdi}. As discussed in Section \ref{flatCon}, the flat space limit implies that for $R^4$ the finite $\tau$ results can be recovered from the 't Hooft limit by simply promoting $\zeta(3)$ to $E(\threeh,\tau_s.\bar\tau_s)$ using \eqref{EisensteinExpansion}, while the difference between $AdS_5\times S^5$ and $AdS_5\times S^5/\mathbb{Z}_2$ theories is only due to the orbifold factor. From the results in \cite{Alday:2018pdi} we can thus obtain
\es{averageR4}{
\langle\lambda^{(0)}_{2,t,\ell}\lambda^{(0)}_{p,t,\ell}\gamma^{R^4}_{t,\ell}\rangle=
\frac{E(\threeh,\tau_s,\bar\tau_s)  (-1)^{p+1} (p)_4 (t -2) t  (t +2)^2 \Gamma \left(\frac{t
   }{2}+4\right)   \Gamma \left(p+\frac{t
   }{2}+2\right) \Gamma\left(\frac{t }{2}+2\right)\delta _{\ell,0}
}{
\orb^\frac{3}{4} \sqrt{2\pi}2^{2 t +12} (p-2)! \Gamma (p+4) \Gamma \left(\frac{t +3}{2}\right) \Gamma \left(\frac{t
   +5}{2}\right)\Gamma\left(\frac t 2-p+1\right)}\,.
}
We can then use this data to perform the $p,t,\ell$ sums for the $\log^2U$ term in \eqref{appA} in a small $z$ expansion as
\es{slices}{
\cG^{A|B}\Big\vert_{\log^2U}=  z^2 h_{A|B}^{(2)}(\bar z) + z^3 h_{A|B}^{(3)}(\bar z) + \cdots \,.
}
The $z$-slices take the same form for orbifold and non-orbifold but differ depending on the vertices $A,B$ as
\es{slices2}{
h^{(n)}_{R|R} (\zb) &= \frac{1}{\zb^{n+2}} \left( p_1^{(n)} (\zb) \text{Li}_2(\zb)
+ p_2^{(n)} (\zb) \log(1-\zb)^2 + p_3^{(n)} (\zb) \log(1-\zb)+ p_4^{(n)} (\zb)\right)\,,\\
h^{(n)}_{A|R^4} (\zb) &= \frac{1}{\zb^{n+3}} \left( p_1^{(n+1)} (\zb) \log(1-\zb)+ p_2^{(n+1)} (\zb)\right)\,,
}
where $A$ can be $R$ or $R^4$, and $p_i^{(n)} (\zb)$ are certain polynomials of degree $n$.

\subsection{Mellin amplitude}
\label{reducedMellinAmplitude}

We can now complete the position space DD to the entire correlator using crossing symmetry in Mellin space following \cite{Alday:2018kkw,Alday:2020tgi}. To convert to Mellin space, we recognize that the $\log^2(z) \log^2(1-\zb)$ terms in \eqref{slices} emerge from simultaneous poles in the Mellin amplitude in $s$ and $t$. Taking residues in \eqref{MellinDef} we see that such poles contribute to the correlator as
\beq
M = \sum\limits_{m,n=2}^\infty \frac{c_{mn}}{(s-2m)(t-2m)} \  \rightarrow \ 
\cT\big|_{\log^2(z) \log^2(1-\zb)} = 
\sum\limits_{m,n=2}^\infty  \frac{c_{mn} \Gamma(m+n)^2}{16 \Gamma(m-1)^2 \Gamma(n-1)^2} U^m V^{n-2}\,.
\label{eq:mellin_4logs}
\eeq
We can now solve for the $c_{mn}$ by comparing \eqref{slices} to \eqref{eq:mellin_4logs} in a double expansion around $z=0$ and $\zb=1$.
After crossing symmetrizing we check that the Mellin amplitude reproduces the full $\log^2(z)$ contribution \eqref{slices}
which implies that there are no further poles in $s$ and, by crossing symmetry, neither in $t$ or $u$.
We  carried out this procedure for each theory and found a similar structure for each case. For $M^{R|R}$ we got
\es{MellinRR}{
&M^{R|R}(s,t) = \sum_{m,n=2}^\infty\Bigg[ \frac{c_{mn}}{(s-2m)(t-2n)} + \frac{c_{mn}}{(t-2m)(u-2n)}+ \frac{c_{mn}}{(u-2m)(s-2n)}-b_{mn}\Bigg]+C\,.
}
For $SU(N)$, the coefficients $c_{mn} = c_{nm}$ were computed in \cite{Alday:2018kkw} as
\begin{align}
c^{SU(N)}_{mn} ={}& 
\frac{(m-1)^2 m^2}{5 (m+n-1)}
+\frac{2(m-1)^2 \left(3 m^2-6 m+8\right)}{5(m+n-2)}
-\frac{9 m^4-54 m^3+123 m^2-126 m+44}{5 (m+n-3)}\nonumber\\
&-\frac{4 \left(m^2-4 m+9\right) (m-2)^2}{5 (m+n-4)}
+\frac{6 (m-3)^2 (m-2)^2}{5 (m+n-5)}\,.
\label{SUNcmn}
\end{align}
These coefficients diverge in the large $m \sim n$ limit, so in \cite{Chester:2019pvm} the $b_{mn}$ was chosen to regulate the sum in this limit, while the constant $C$ was then chosen so that the Mellin amplitude matches the position space correlator of \cite{Aprile:2017bgs}, which gave
\es{bC}{
b^{SU(N)}_{mn} = \frac{9mn}{2(m+n)^3}\,, \qquad
C^{SU(N)} = -\frac{39}{16} - \frac{13}{8} \pi^2 + 9 \zeta(3)\,.
}
As shown in \cite{Aprile:2017bgs}, this convention for the 1-loop Mellin amplitude gives CFT data that is analytic in spin including spin zero, while other conventions would change the constant contact term and so would give different results for spin zero. Since we anyway add the contact term ambiguity $\bar B_0^{R|R}$ in \eqref{MIntro}, the value of $C$ here is a matter of convenience. In Appendix \ref{M_RR_App} we explain how this Mellin amplitude can be summed to give a closed form expression.

For the orbifold cases, we found that the $c_{mn}$ are related to those of $SU(N)$ as
\es{SONcmn}{
c^{orb}_{mn} = \frac12 c^{SU(N)}_{mn} + d_{mn}\,,
}
where the new term $d_{mn}$ takes the form
\es{d}{
d_{mn} ={}& \frac{d^{(1)}_{m n,m+n-1} \, (m+n-1)  +  d^{(2)}_{m n,m+n-1} \, {}_3 F_2 \left( 1,1-m,1-n; \frac32, 2-m-n; 1\right)}{768 (m-1)^2 (2 m+1) (n-1)^2 (2 n+1) (m+n-1)}\,,
}
and we define the polynomials
\es{dpoly}{
d^{(1)}_{p,q}={}&
-32 p^5 \left(q^2-5 q+2\right)+16 p^4 \left(23 q^3-80 q^2-45 q+14\right)\\
&-16 p^3 \left(91 q^4-256 q^3-173 q^2+34
   q-18\right)\\
&+8 p^2 \left(269 q^5-801 q^4+79 q^3+307 q^2+18\right)\\
&-2 p q^2 \left(201 q^4-1731 q^3+3841 q^2+1323
   q+414\right)\\
&+q^2 \left(-945 q^5+1050 q^4+3842 q^3+852 q^2+783 q-270\right)\,,\\
d^{(2)}_{p,q}={}&
64 p^6 \left(q^2-5 q+2\right)-128 p^5 \left(6 q^3-21 q^2-12 q+4\right)\\
&+16 p^4 \left(205 q^4-570 q^3-467 q^2+40
   q-24\right)\\
&-64 p^3 q \left(90 q^4-244 q^3-81 q^2+53 q+6\right)\\
&+4 p^2 q \left(739 q^5-2917 q^4+2693 q^3+2361
   q^2+840 q-36\right)\\
&+8 p q^3 \left(186 q^4+163 q^3-1678 q^2-1251 q-540\right)\\
&+3 q^3 \left(-315 q^5+140 q^4+1500
   q^3+1162 q^2+495 q+90\right)\,.
}
In order to obtain this result it was useful to recognize that for 
different fixed values of $m$, the coefficients can be written as
\beq
c^{orb}_{mn} = \sum\limits_{j=0}^{m} \frac{\mu_{m,j}}{m+n-1-j}\,.
\eeq
We were then able to determine a general formula for $\mu_{m,j}$.
Unlike $c^{SU(N)}_{mn} $, the sum over the $d_{mn}$ is convergent, so we can regularize the orbifold case similar to $SU(N)$ as
\es{bC2}{
b^{orb}_{mn} = \frac12 b^{SU(N)}_{mn}\,, \qquad
C^{orb} = \frac12C^{SU(N)}\,,
}
where the value of $C^{orb}$ is again a choice, that we will find in the next section also leads to CFT data analytic in spin down to spin zero.

For $M^{R|R^4}(s,t)$ and $M^{R^4|R^4}(s,t)$ we similarly convert the position space DD to Mellin space to find the simpler structure
\es{RR4c}{
M^{A|R^4}(s,t) &= \sum_{m=2}^\infty\left( \frac{ c_{m}}{(s-2m)} +  \frac{ c_{m}}{(t-2m)} +  \frac{ c_{m}}{(u-2m)}  \right)+\text{polynomial}\,,
}
where $A$ will be $R$ or $R^4$, and the polynomial (which we omit in the results below) is any crossing symmetric polynomial of degree four for $M^{R|R^4}(s,t)$ and degree seven for $M^{R^4|R^4}(s,t)$, since these are the degrees determined by the powers of $c$ and the flat space limit. The results for $SU(N)$ theories were given in \cite{Alday:2018kkw} in the 't Hooft limit\footnote{Up to a few typos, that were corrected in \cite{Drummond:2019hel}.}, while in the finite $\tau$ limit they just differ by an overall factor:
\es{RR4SU(N)}{
c^{R|R^4,SU(N)}_{m} ={}& -\frac{ E(\threeh,\tau,\bar\tau)}{2\sqrt{2}\pi^{\frac32}} \left(63 m^4-322 m^3+693 m^2-722 m+300\right)\,,\\
c^{R^4|R^4,SU(N)}_{m} ={}& -\frac{135E(\threeh,\tau,\bar\tau)^2 }{896\pi^3}  \big(924 m^7-11627 m^6+67137 m^5-227045 m^4\\
&+480151 m^3-629468 m^2+470408 m-153720\big)\,.
}
For the orbifold case, we found that the coefficients are related to $SU(N)$ similar to the $R|R$ case as
\es{RR4SO(N)}{
c^{R|R^4,orb}_{m} = \frac{1}{2^\frac{3}{4} 2} c^{R|R^4,SU(N)}_{m} + d^{R|R^4}_{m}\,,\qquad
c^{R^4|R^4,orb}_{m} = \frac{1}{2^{\frac32} 2} c^{R^4|R^4,SU(N)}_{m} + d^{R^4|R^4}_{m}\,,
}
where the new terms are defined as
\begin{align}
d^{R|R^4}_{m} ={}& \frac{E(\threeh,\tau,\bar\tau)}{64\cdot 2^\frac{1}{4} \pi^\frac32} \left(\frac{3^3 5^3 \left( -17488 m^3+10854 m^2+6031 m-1506 \right)}{2^{11} \left(m-\frac{15}{2} \right)_9} - \frac{15}{4} (3m-2) \right)\,,\nonumber\\
d^{R^4|R^4}_{m} ={}& \frac{E(\threeh,\tau,\bar\tau)^2}{\sqrt{2} \pi^3} \frac{3^6 5^4 }{2^{22}\left(m-\frac{21}{2} \right)_{13}} \Big( 1062890 m^7-1606305 m^6-1152985 m^5\nonumber\\
&+3799305 m^4 -2802325 m^3+186276 \Big)\,,
\label{dR4}
\end{align}
and note the orbifold factor $2^{\frac34}$ due to the $R^4$ vertex.

The sum over $m$ in \eqref{RR4c} for these coefficients is divergent, but can be regulated and resummed as described in \cite{Alday:2018kkw} to get for $SU(N)$ 
\begin{align}
M^{R|R^4}_{SU(N)}=& -\frac{ E(\threeh,\tau,\bar\tau)}{64\sqrt{2}\pi^{\frac32}} \left(63 s^4-644 s^3+2772 s^2-5776 s+4800\right)  \psi\left(2-\frac{s}{2}\right)+\text{crossed}\,,\nonumber\\
M^{R^4|R^4}_{SU(N)} =& -\frac{135E(\threeh,\tau,\bar\tau)^2}{2^{14}\cdot 7\pi^3} \Big(462 s^7-11627 s^6+134274 s^5-908180 s^4+3841208 s^3\nonumber\\
&-10071488 s^2+15053056 s-9838080\Big)
    \psi\left(2-\frac{s}{2}\right)+\text{crossed}\,,
\label{RR4finalSUN}
\end{align}
and for the orbifold case

\begin{align}
&M^{R|R^4}_{orb} = \frac{1}{2^\frac{3}{4} 2} M^{R|R^4}_{SU(N)}  + \frac{E(\threeh,\tau,\bar\tau)}{64\cdot2^{\frac{1}{4}}\pi^{\frac32}} \bigg(
\frac{15}{16} (4-3 s) \psi\left(2-\frac{s}{2}\right)\nonumber\\
&+ \frac{3^2 5^2}{2^{17}} \bigg(
\frac{9 n_{-1}(s)}{s+1}-\frac{388 n_1(s)}{s-1}-\frac{18040 n_3(s)}{s-3}+\frac{255788 n_5(s)}{s-5}-\frac{995390
   n_7(s)}{s-7}+\frac{1797556 n_9(s)}{s-9}\nonumber\\
&-\frac{1699712 n_{11}(s)}{s-11}+\frac{820260
   n_{13}(s)}{s-13}-\frac{160083 n_{15}(s)}{s-15}
 \bigg)\bigg) + \, \text{crossed}  \,, \nonumber\\
&M^{R^4|R^4}_{orb} = \frac{1}{2^{\frac32} 2} M^{R^4|R^4}_{SU(N)} + \frac{E(\threeh,\tau,\bar\tau)^2}{4096\sqrt{2}\pi^3} \frac{3^5 5^2}{2^{25}} \bigg(
-\frac{3 n_{-3}(s)}{s+3}+\frac{80 n_{-1}(s)}{s+1}+\frac{4790 n_1(s)}{s-1}\nonumber\\
&-\frac{2492440 n_3(s)}{s-3}+\frac{638579095 n_5(s)}{s-5}-\frac{15176687744 n_7(s)}{s-7}+\frac{121176047620 n_9(s)}{s-9}
\label{RR4finalSON}\\
&-\frac{465561453840 n_{11}(s)}{s-11}+\frac{996963105915 n_{13}(s)}{s-13}-\frac{1260126418320
   n_{15}(s)}{s-15}\nonumber\\
&+\frac{937494179622 n_{17}(s)}{s-17}-\frac{380559979800 n_{19}(s)}{s-19}+\frac{65155115025
   n_{21}(s)}{s-21}
\bigg) + \, \text{crossed}\,,
\nonumber
\end{align}
where the functions
\es{n}{
n_k(s) \equiv \psi\left(2-\frac{s}{2}\right) - \psi\left(2-\frac{k}{2}\right)\,,
}
have a zero at $s=k$ which cancels all the apparent poles at odd integers.

\subsection{Flat space limit}
\label{flatsec}

We will now compare the 1-loop Mellin amplitudes to the corresponding string theory amplitudes in 10d using the flat-space limit formula \eqref{flat}. To apply this formula to the 1-loop amplitudes, we should look at the regime where $m,n,s,t,u$ all scale equally large.
One can check that the terms $d_{mn}$ and $d_m$ in \eqref{SONcmn} and \eqref{RR4SO(N)} are subleading in this limit.
Hence the leading terms of each 1-loop Mellin amplitude in \eqref{SONcmn} and \eqref{RR4SO(N)} differ for $SU(N)$ and the orbifold theories by the same powers of 2 that appear in the respective AdS/CFT dictionaries \eqref{cPlanck} and it is only necessary to check $SU(N)$. The $R|R$ term was already checked in \cite{Alday:2018kkw}, so it suffices to check $R|R^4$ and $R^4|R^4$. From \eqref{RR4finalSUN} we find that
\es{flatM}{
f_{SU(N)}^{R|R^4}(s,t) &= -\frac{\pi ^2E(\threeh,\tau,\bar\tau)}{2^{12} 45}  g_s^{\frac72} \ell_s^{14} s^4 \log (s) + \, \text{crossed}\,, \\
f_{SU(N)}^{R^4|R^4}(s,t) &= -\frac{\pi ^2E(\threeh,\tau,\bar\tau)^2}{2^{20} 105}  g_s^5 \ell_s^{20} s^7 \log (s) + \, \text{crossed}\,.
}
We can compare this to the discontinuity of the respective 10d amplitude, as computed from the unitarity cut formula \cite{Green:2008uj}\footnote{We fix various typos in \cite{Green:2008uj} and \cite{Alday:2018pdi}, by comparing our formula to the 10d box diagram in e.g. \cite{Alday:2017vkk}.}
\es{unitaritycut}{
 {\text{Disc}_s f^{A|B} (s,t)} =&(\delta_{A,B}-2) s t u \frac{\pi^2 i g_s^2 l_s^8 s^{7}}{15\, 2^{9}}\int_0^\pi d\theta\int_0^{2\pi} d\phi \sin^7\theta |\sin^6\phi| \frac{f^A(s,t')}{s t' u' } \frac{f^B(s,t'')}{s t'' u''}\,,
}
with
\beq
t'=-\frac{s}{2}(1-\cos\theta)\,,\quad  t''=-\frac s2(1+\cos\theta\cos\rho+\sin\theta\cos\phi\sin\rho)\,,
\quad u'=-s-t'\,, \quad u'' = -s-t''\,,
\eeq
and $\cos \rho = \frac{t-u}{s}$.
Using the vertices in \eqref{fEisenstein}, we find a precise match with \eqref{flatM}.

\subsection{Fixing 1-loop contact term from localization}
\label{contact1loop}

We can also use the explicit $R|R$ Mellin amplitudes and the localization constraint in \eqref{constraint1} to fix the contact term ambiguity $\bar B_0^{R|R}$. In \cite{Chester:2019pvm}, this was done for the $SU(N)$ case by numerically computing the integral of the explicit position space amplitude in \cite{Aprile:2017bgs}. Here, we will fix both $SU(N)$ and the orbifold theories by computing the Mellin space integral in \eqref{Mints} analytically. The calculation is described in Appendix \ref{locInt}, and the result is
\es{1loopInts}{
I_2[M_{SU(N)}^{R|R}]=\frac{5}{32}\,,\qquad I_2[M_{orb}^{R|R}]=\frac{1}{128}\,.
}
We can then input the localization inputs \eqref{tohooftSUN} and \eqref{tohooftSON} to the constraint \eqref{constraint1} to fix $\bar B_0^{R|R}$ to 
\es{Bfix1loop}{
\bar B_{0,SU(N)}^{R|R}=\frac{15}{4}\,,\qquad \bar B_{0,orb}^{R|R}=\frac{15}{8}\,.
} 
Curiously, these terms differ by the same factor of two as the leading $s,t$ term in $M^{R|R}$ that contributes to the flat space limit, even though these contact terms are subleading in the AdS radius and so do not contribute to the flat space limit.

\subsection{Extracting CFT data}
\label{CFTData}

Lastly, we can extract all low-lying CFT data from the $R|R$, i.e. $c^{-2}$, correlator using two methods. Firstly, in Appendix \ref{inversion} we derive an inversion integral formula for each DD in position space following \cite{Alday:2017vkk}, which allows us to efficiently extract all CFT data above spin $\ell=0$, as expected from the Lorentzian inversion formula \cite{Caron-Huot:2017vep}. Secondly, we use the projection method of \cite{Chester:2018lbz} to expand each correlator as written in Mellin space in conformal blocks and extract the CFT data for all spins including $\ell=0$, using that the contact term ambiguity $\bar B_0^{R|R}$ was fixed in the previous section. Not only does each method agree for $\ell>0$, but we even find that the analytic continuation of the inversion method to spin zero matches the contribution from $M^{R|R}$, i.e. neglecting $\bar B_0^{R|R}$, for both $SU(N)$ and the orbifold theories. We can also use the projection method to compute the contribution from the tree level terms that we fixed in Section \ref{locCon}. We do not extract CFT data from the $R|R^4$ or $R^4|R^4$ correlators, since for low spins these terms are affected by contact terms for which we have insufficient localization constraints to fix them.

Combining these various contributions, for the lowest dimension operators for each spin $\ell$ that appear in the large $c$ and finite $\tau$ limit (which are all double trace operators),\footnote{These are the operators introduced in Section \ref{1loopfrom} with $p=2, n=0$.} we find for the lowest few $\ell$:
\begin{align}
\Delta_{0,SU(N)}^{R|R}={}&4-\frac4c-\frac{135
   E(\threeh, \tau, \bar \tau)}{7 \sqrt{2} \pi ^{3/2}c^{\frac74}}+\frac{1199}{42 c^2}-\frac{3825  E(\fiveh, \tau, \bar \tau)}{32 \sqrt{2} \pi ^{5/2}c^{\frac94}}+O(c^{-\frac52})\,,\nonumber\\
\Delta_{2,SU(N)}^{R|R}={}&6-\frac1c-\frac{41}{16 c^2}-\frac{1575 E(\fiveh, \tau, \bar \tau)}{22 \sqrt{2} \pi ^{5/2}c^{\frac94}}+O(c^{-\frac52})\,,\nonumber\\
\Delta_{4,SU(N)}^{R|R}={}&8-\frac{12}{25 c}-\frac{423}{3125 c^2}+O(c^{-\frac52})\,,\nonumber\\
\Delta_{0,orb}^{R|R}={}&4-\frac{4}{c}-\frac{135 E(\threeh, \tau_s, \bar \tau_s)}{14
   \sqrt[4]{2} \pi ^{3/2}c^{\frac74}}+\frac{1}{c^2} \left(\tfrac{29129625 \zeta (3)+22143194}{7680}-\tfrac{135}{14}-10688 \log (2)\right)\nonumber\\
&-\frac{3825 
   E(\fiveh, \tau_s, \bar \tau_s)}{64\ 2^{3/4} \pi ^{5/2}c^{\frac94}}+O(c^{-\frac52})\,,
\label{anoms}\\
\Delta_{2,orb}^{R|R}={}&6-\frac{1}{c}+\frac{1}{c^2}\left(\tfrac{469744545375 \zeta (3)+353499406166}{1720320}-769988 \log (2)\right)\nonumber\\
&-\frac{1575
   E(\fiveh, \tau_s, \bar \tau_s)}{44\ 2^{3/4} \pi ^{5/2}c^{\frac94}}+O(c^{-\frac52})\,,
\nonumber\\
\Delta_{4,orb}^{R|R}={}&8-\frac{12}{25 c}-\frac{9}{c^2} \left( -\tfrac{217777351113 \zeta
   (3)}{409600}-\tfrac{71700177905311}{179200000}+\tfrac{937053696 \log (2)}{625} \right)+O(c^{-\frac52})\,.\nonumber
\end{align}
One could similarly extract the OPE coefficients for these lowest twist operators, but those are harder to bound using the numerical bootstrap,\footnote{This is because in order to bound OPE coefficients of unprotected operators, one must first know their scaling dimensions, for which there is extra uncertainty from their own bounds.} so we leave it for future work. For higher twist operators, due to the degeneracy discussed above, one would need to look at more correlators to perform their unmixing at 1-loop order.

\section{Comparison to numerical bootstrap}
\label{numerics}

\begin{figure}[]
\begin{center}
       \includegraphics[width=.63\textwidth]{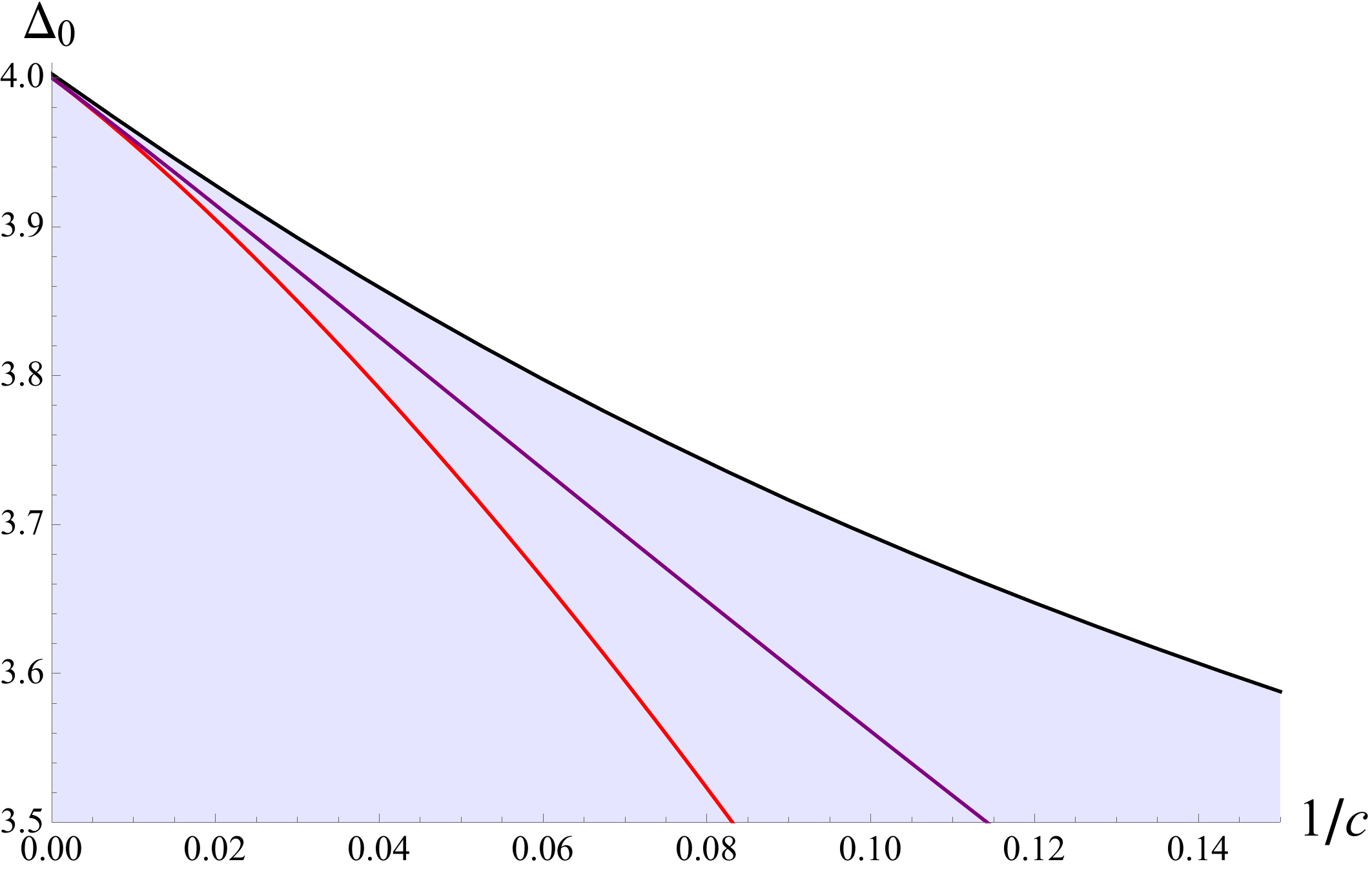}
      \includegraphics[width=.63\textwidth]{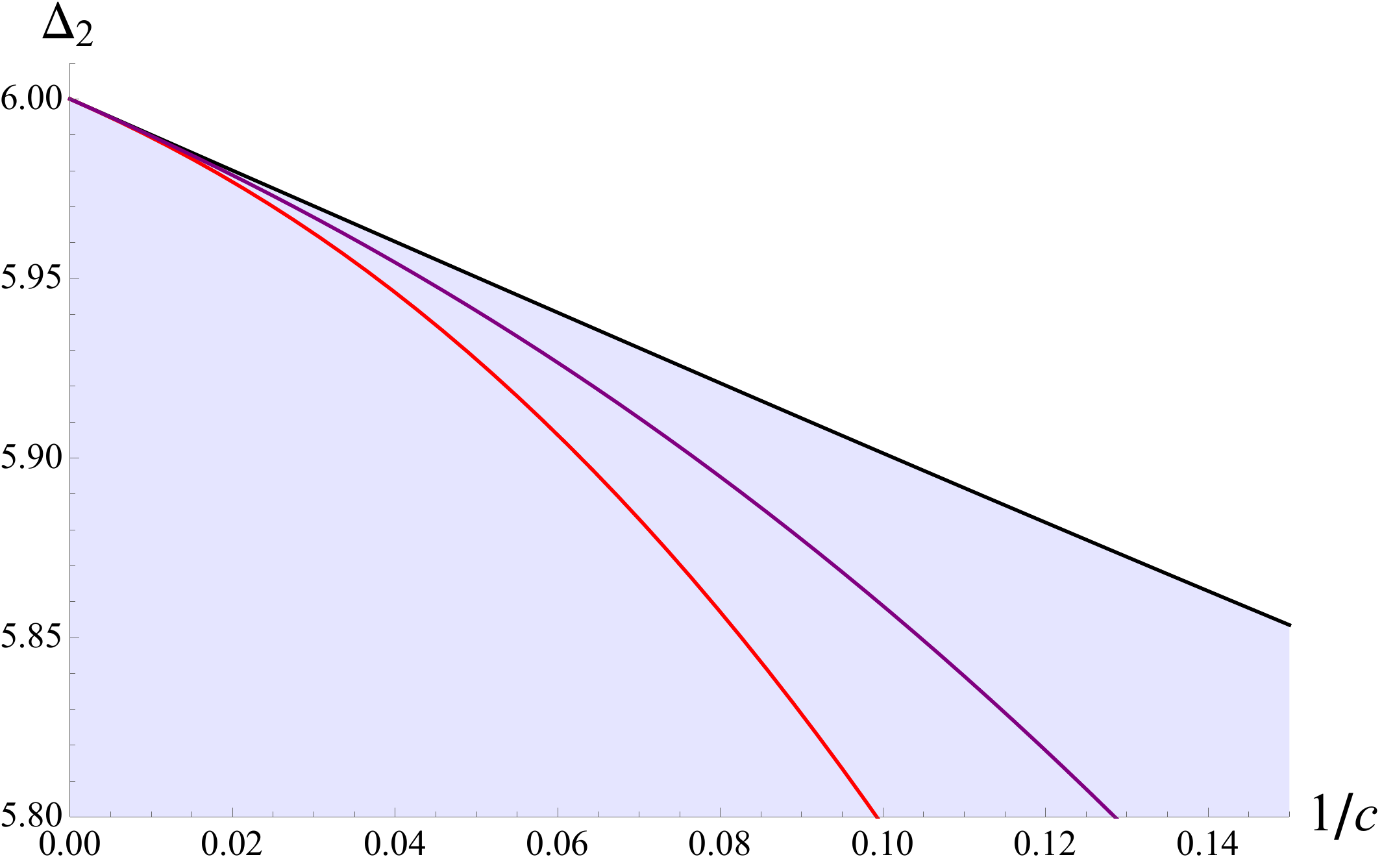}
            \includegraphics[width=.63\textwidth]{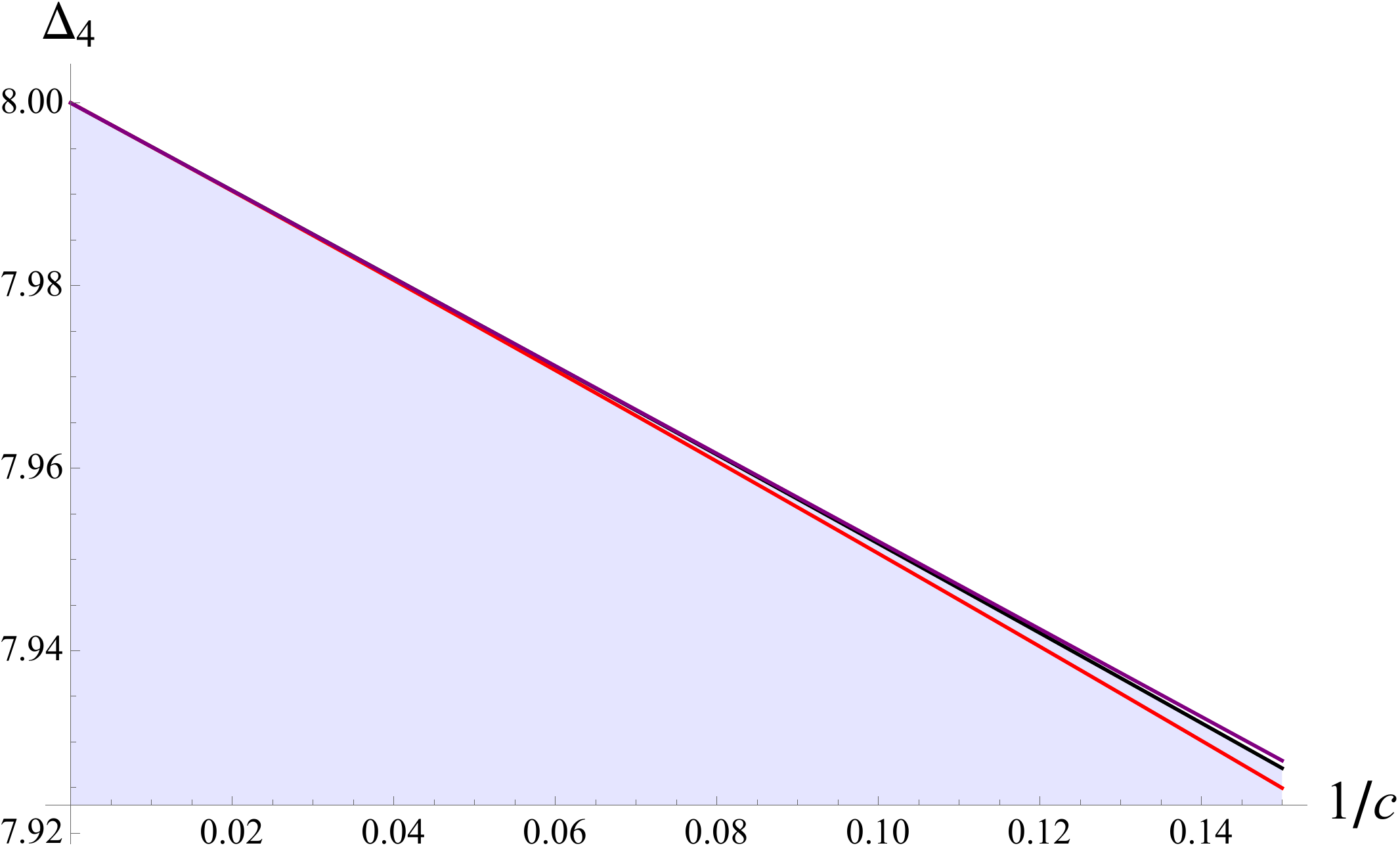}
\caption{Upper bounds (black) on the lowest dimension scaling dimension for spin $0,2,4$ in terms of $c$, computed using $n_\text{max}=62$ for the single correlator setup, or $n_\text{max}=30$ for the mixed correlator setup (the results cannot be distinguished by eye in this regime of large $c$). The red, purple lines corresponds to the $SU(N)$, $SO(2N)$ analytic estimates, respectively, at large $c$ and at the self dual point $\tau=e^{i\pi/3}$, as given in \eqref{anomsNum}.}
\label{bootBounds}
\end{center}
\end{figure}

We conclude by comparing our large $c$ results to finite $c$ numerical bootstrap bounds. In \cite{Beem:2013qxa,Beem:2016wfs}, bounds were computed on CFT data that appears in the stress tensor correlator as a function of $c$. We restrict the discussion in this section to the $SU(N)$ or $SO(2N)$ groups, since at large $c$ and finite $\tau$ all the other orbifold gauge theories can be easily related to $SO(2N)$. Since the value of $\tau$ could not be specified \cite{Beem:2013qxa,Beem:2016wfs}, it was conjectured that the bounds might be saturated by either the $SU(N)$ or $SO(2N)$ theories at one of the self-dual points under S-duality, which are either $\tau=i$ with $\mathbb{Z}_2$ enhancement or $\tau=e^{i\pi/3}$ with $\mathbb{Z}_3$ enhancement. We can compute the anomalous dimensions \eqref{anoms} at the self-dual points using the formulae \cite{Alday:2013bha}:
\es{alday}{
E(s, e^{\frac{i\pi}{3}})&=2^{1-s} 3^{1-\frac{s}{2}} \zeta (s) \left(\zeta \left(s,\frac{1}{3}\right)-\zeta \left(s,\frac{2}{3}\right)\right)\,,\\
E(s,i)&=4^{1-s} \zeta (s) \left(\zeta \left(s,\frac{1}{4}\right)-\zeta \left(s,\frac{3}{4}\right)\right)\,,
}
which shows that $\tau=e^{i\pi/3}$ has the largest value, which numerically is
\es{anomsNum}{
\tau=e^{i\pi/3}:\qquad \Delta_{0,SU(N)}^{R|R}&=4-\frac4c-\frac{21.7787}{c^{\frac74}} +\frac{28.5476}{c^2}-\frac{22.8027}{c^{\frac94}} +O(c^{-\frac52})\,,\\
\Delta_{2,SU(N)}^{R|R}&=6-\frac1c-\frac{2.5625}{c^2}-\frac{13.6573}{c^{\frac94}}+O(c^{-\frac52}) \,,\\
\Delta_{4,SU(N)}^{R|R}&=8 -\frac{0.48}{c}-\frac{0.13536}{c^2}+O(c^{-\frac52})\,,\\
\Delta_{0,SO(2N)}^{R|R}&=4-\frac4c-\frac{12.9497}{c^{\frac74}} +\frac{24.534}{c^2}-\frac{9.58737}{c^{\frac94}}+O(c^{-\frac52})\,,\\
\Delta_{2,SO(2N)}^{R|R}&=6-\frac1c-\frac{0.888511}{c^2}-\frac{5.74217}{c^{\frac94}} +O(c^{-\frac52})\,,\\
\Delta_{4,SO(2N)}^{R|R}&=8 -\frac{0.48}{c}-\frac{0.00174726}{c^2}+O(c^{-\frac52})\,.\\
}
The $SO(2N)$ value is also larger than the $SU(N)$ value for large $c$, so if any theory saturates the single correlator bounds of \cite{Beem:2013qxa,Beem:2016wfs}, it should be the $SO(2N)$ theory at $\tau=e^{i\pi/3}$. A numerical bootstrap study was also done in \cite{Bissi:2020jve} using mixed correlators between the stress tensor multiplet and the next lowest half-BPS multiplet with $\Delta = 3$. Since this multiplet is absent for $SO(2N)$, only the $SU(N)$ theory could appear in these bounds.
In Figure \ref{bootBounds} we show a comparison between the single and mixed correlator bounds which we recomputed at very high bootstrap precision for low spins, and the analytic estimates at large $c$.\footnote{We include the explicit data points in the attached \texttt{Mathematica} file.}
For completeness we give a full description of our mixed correlator bootstrap setup (independent of \cite{Bissi:2020jve}) in Appendix \ref{mixedCorr}.
None of the analytic estimates for $\Delta_0$ and $\Delta_2$ are close to saturating the bounds,\footnote{Truncating the analytic expansion to a smaller order in large $c$ also does not help.} and the single correlator and mixed correlator bounds are in fact almost indistinguishable in this regime, which suggests that no physical theory is saturating these bounds.

\section*{Acknowledgments} 

We thank Ofer Aharony, Silvu Pufu, Yifan Wang, and Congkao Wen for useful conversations. The work of LFA and TH is supported by the European Research Council (ERC) under the European Union's Horizon
2020 research and innovation programme (grant agreement No 787185). LFA is also supported in part by the STFC grant ST/T000864/1. SMC is supported by the Zuckerman STEM Leadership Fellowship. The authors would like to acknowledge the use of the University of Oxford Advanced Research Computing (ARC) facility in carrying out this work.(http://dx.doi.org/10.5281/zenodo.22558)
\appendix

\section{Matrix model calculations}
\label{locApp}

In this appendix we give details of the matrix model computation in Section \ref{finiteN}. As discussed in \cite{Chester:2019pvm}, the mass derivatives that we consider in \eqref{constraint1} are identical for $SU(N)$ and $U(N)$, so for simplicity we will consider the latter gauge group. For the various gauge groups we consider, we will use the standard basis of roots given in Appendix A of \cite{Fiol:2014fla} to write \eqref{Z} explicitly as
\begin{align}
& U(N):\;\; Z(m)=\int\frac{ d^Na}{N!} \frac{e^{-\frac{8\pi^2}{g_\text{YM}^2}\sum_ia_i^2}}{H(m)^N}  \prod_{i<j}\frac{a_{ij}^2H^2(a_{ij})}{H(a_{ij}-m)H(a_{ij}+m)}\,,\nonumber\\
& SO(2N):\;\;  Z(m)=\int \frac{d^Na}{2^NN!} \frac{e^{-\frac{8\pi^2}{g_\text{YM}^2}\sum_ia_i^2}}{H(m)^N} \prod_{i<j}\frac{a_{ij}^2(a_{ij}^+)^2H^2(a_{ij})H^2(a_{ij}^+)}{H(a_{ij}-m)H(a_{ij}+m)H(a_{ij}^+-m)H(a_{ij}^++m)}\,,\nonumber\\
 & SO(2N+1):\;\;  Z(m)=\int \frac{d^Na}{2^NN!} \frac{e^{-\frac{8\pi^2}{(\delta_{N,1}+1)g_\text{YM}^2}\sum_ia_i^2}}{H(m)^N}\prod_i\frac{a_i^2H^2(a_i)}{H(a_i+m)H(a_i-m)}
\label{ZUN}\\
 &\qquad \qquad \qquad \quad \times \prod_{i<j}\frac{a_{ij}^2(a_{ij}^+)^2H^2(a_{ij})H^2(a_{ij}^+)}{H(a_{ij}-m)H(a_{ij}+m)H(a_{ij}^+-m)H(a_{ij}^++m)}\,,\nonumber\\
  & USp(2N):\;\;  Z(m)=\int \frac{d^Na}{2^NN!} \frac{e^{-\frac{16\pi^2}{g_\text{YM}^2}\sum_ia_i^2}}{H(m)^N}\prod_i\frac{(2a_i)^2H^2(2a_i)}{H(2a_i+m)H(2a_i-m)}
\nonumber\\
 &\qquad \qquad \qquad \quad \times \prod_{i<j}\frac{a_{ij}^2(a_{ij}^+)^2H^2(a_{ij})H^2(a_{ij}^+)}{H(a_{ij}-m)H(a_{ij}+m)H(a_{ij}^+-m)H(a_{ij}^++m)}\,,\nonumber
\end{align}
 where we define $a_{ij}=a_i-a_j$ and $a_{ij}^+=a_i+a_j$, and note that the Killing form for $USp(2N)$ in our basis has extra factors of $2$ relative to the other cases, while $SO(3)$ has an extra factor of $\frac12$. We will compute the small $m$ expansion of $Z(m)$ using the method of orthogonal polynomials.

Let us introduce a family of polynomials $p_n(a)$:
 \es{pn}{
 p_n(a)\equiv \left(\frac{g_\text{YM}^2}{32\pi^2 }\right)^{\frac n2}H_n\left(\frac{4\pi a}{\sqrt{2}g_\text{YM}}\right)\,,
 }
 which are orthogonal with respect to the Gaussian matrix model measure:
 \es{ortho}{
 \int da p_m(a)p_n(a)e^{-\frac{8\pi^2}{g_\text{YM}^2}a^2}=n!\left(\frac{g_\text{YM}^2}{16\pi^2 }\right)^n\sqrt{\frac{g_\text{YM}^2}{8\pi }}\delta_{mn}\equiv h_n\delta_{mn}\,.
 }
These orthogonal polynomials are useful because they diagonalize the Vandermonde determinant for the various Gaussian matrix models we consider:
\es{gaussUN}{
U(N):\;\;Z(0)&=\frac{1}{N!}\int d^Na\prod_{i<j}{(a_i-a_j)^2}e^{-\frac{8\pi^2}{g_\text{YM}^2}\sum_ia_i^2}
\\
&=\frac{1}{N!}\int d^Na\prod_{i<j}|p_{i-1}(a_j)|^2e^{-\frac{8\pi^2}{g_\text{YM}^2}\sum_ia_i^2}\\
&=\prod_{k=0}^{N-1}h_k\,,\\
&=2^{\frac{1}{2} (1-4 N) N} \pi
   ^{\frac{1}{2} (1-2 N) N} G(N+1) (g_\text{YM}^2)
   ^{\frac{N^2}{2}}\,,
}
and
\es{gaussSON}{
SO(2N):\;\;Z(0)&=\frac{1}{2^NN!}\int d^Na\prod_{i<j}{(a^2_i-a^2_j)^2}e^{-\frac{8\pi^2}{g_\text{YM}^2}\sum_ia_i^2}\\
&=\frac{1}{2^NN!}\int d^Na\prod_{i<j}|p_{2(i-1)}(a_j)|^2e^{-\frac{8\pi^2}{g_\text{YM}^2}\sum_ia_i^2}\\
&=\frac{1}{2^N}\prod_{k=0}^{N-1}h_{2k}\,,\\
&=\frac{A^{3/2} 2^{-3 N^2+\frac{N}{2}-\frac{1}{24}} \pi ^{-2 N^2+N+\frac{1}{4}}
   G\left(N+\frac{1}{2}\right) G(N+1) (g_\text{YM}^2)
   ^{N^2-\frac{N}{2}}}{\sqrt[8]{e}}\,,\\
}
and
\begin{align}
SO(2N+1):\;\;Z(0)&=\frac{1}{2^NN!}\int d^Na\prod_{i<j}{(a^2_i-a^2_j)^2}\prod_ia_i^2e^{-\frac{8\pi^2}{g_\text{YM}^2}\sum_ia_i^2}\nonumber\\
&=\frac{1}{2^NN!}\int d^Na\prod_{i<j}|p_{2i-1}(a_j)|^2e^{-\frac{8\pi^2}{g_\text{YM}^2}\sum_ia_i^2}\nonumber\\
&=\frac{1}{2^N}\prod_{k=0}^{N-1}h_{2k}\,,
\label{gaussSON2}\\
&=\frac{A^{3/2} 2^{-3N^2-\frac{5N}{2}-\frac{1}{24}}
 \pi ^{-2
  N^2-N-\frac{1}{4}} G(N+1) G\left(N+\frac{3}{2}\right)
   (g_\text{YM}^2) ^{N^2+\frac{N}{2}}}{\sqrt[8]{e}}\,,
\nonumber
\end{align}
where $A$ is the Glaisher constant, $G(z)$ is the Barnes G function, and for $SO(2N+1)$ we assume $N>1$ so we have the standard Gaussian measure. Here, we replaced the Vandermonde determinant by a determinant of orthogonal polynomials that we can write as
 \es{orthoDet}{
\prod_{i<j}|p_{i-1}(a_j)|^2= \sum_{\sigma_1\in S_N}(-1)^{|\sigma_1|}\prod_{k_1=1}^N p_{\sigma_1(k_1)-1}(a_{k_1})\sum_{\sigma_2\in S_N}(-1)^{|\sigma_2|}\prod_{k_2=1}^Np_{\sigma_2(k_2)-1}(a_{k_2})\,,
 }
  and then we computed each integral using \eqref{ortho}. We can similarly compute $USp(2N)$, except we now set $g^2_\text{YM}\to g^2_\text{YM}/2$ in the polynomials $p_n(a)$ to account for the different Gaussian factor, which gives
\begin{align}
USp(2N):\;\;Z(0)&=\frac{1}{2^NN!}\int d^Na\prod_{i<j}{(a^2_i-a^2_j)^2}\prod_i(2a_i)^2e^{-\frac{16\pi^2}{g_\text{YM}^2}\sum_ia_i^2}\\
&=2^N \frac{A^{3/2} 2^{-3N^2-\frac{5N}{2}-\frac{1}{24}}
 \pi ^{-2
  N^2-N-\frac{1}{4}} G(N+1) G\left(N+\frac{3}{2}\right)
   (g_\text{YM}^2/2) ^{N^2+\frac{N}{2}}}{\sqrt[8]{e}}\,.\nonumber
\end{align}
 We can then take the logarithm to get the results in \eqref{Fnom}, where note that $c=N^2/4$ for $U(N)$. For $SU(N)$, we would get $c=(N^2-1)/4$ because there is one less eigenvalue, as discussed in \cite{Chester:2019pvm}.
  
Next, we will discuss how to compute the expectation values in \eqref{2m2L}.  Consider an $n$-body operator $\cO_n(a)$ that WLOG only depends on the $a_i$ for $i=1,\dots n$. Due to the orthogonality of the polynomials in $\langle\cO_n(a)\rangle$ the only permutations $\sigma_1,\sigma_2$ in \eqref{orthoDet} that survive integration are those for which $\sigma_2(m)=\sigma_1(m)$ for $m>n$. This means that in order to contribute to the full matrix model integral, $\{\sigma_2(1),\dots,\sigma_2(n)\}$ must be a permutation of $\{\sigma_1(1),\dots,\sigma_1(n)\}$, which we denote by $\mu$. The expectation value for each gauge group is
 \es{nbodyExp}{
&U(N):\;\; \langle\cO_n(a)\rangle=\frac{1}{N!}\sum_{\substack{\sigma\in S_N\\\mu\in S_n}}(-1)^{|\mu|}\int\left(\prod_{i=1}^n da_i\frac{p_{\sigma(i)-1}(a_i)p_{\mu(\sigma(i))-1}(a_i)}{h_{\sigma(i)-1}}e^{-\frac{8\pi^2}{g_\text{YM}^2}a_i^2}\right)\cO_n(a)\,,\\
&SO(2N):\;\; \langle\cO_n(a)\rangle=\frac{1}{N!}\sum_{\substack{\sigma\in S_N\\\mu\in S_n}}(-1)^{|\mu|}\int\left(\prod_{i=1}^n da_i\frac{p_{2(\sigma(i)-1)}(a_i)p_{2(\mu(\sigma(i))-1)}(a_i)}{h_{2(\sigma(i)-1)}}e^{-\frac{8\pi^2}{g_\text{YM}^2}a_i^2}\right)\cO_n(a)\,,\\
&SO(2N+1) :\;\; \langle\cO_n(a)\rangle=\frac{1}{N!}\sum_{\substack{\sigma\in S_N\\\mu\in S_n}}(-1)^{|\mu|}\int\left(\prod_{i=1}^n da_i\frac{p_{2\sigma(i)-1}(a_i)p_{2\mu(\sigma(i))-1}(a_i)}{h_{2\sigma(i)-1}}e^{-\frac{8\pi^2}{g_\text{YM}^2}a_i^2}\right)\cO_n(a)\,,\\
&USp(2N) :\;\; \langle\cO_n(a)\rangle=\frac{1}{N!}\sum_{\substack{\sigma\in S_N\\\mu\in S_n}}(-1)^{|\mu|}\int\left(\prod_{i=1}^n da_i\frac{p_{2\sigma(i)-1}(a_i)p_{2\mu(\sigma(i))-1}(a_i)}{h_{2\sigma(i)-1}}e^{-\frac{16\pi^2}{g_\text{YM}^2}a_i^2}\right)\cO_n(a)\,,\\
 }
 where note that the originally $N$-dimensional integral has reduced to an $n$-dimensional integral, and the $p_n(a)$ in the $USp(2N)$ case are defined with $g^2_\text{YM}\to g^2_\text{YM}/2$ as usual for this case. We can now use this and the identity
 \es{identity}{
 \int_{-\infty}^\infty e^{-x^2+yx}H_m(x)H_n(x)=e^{\frac{y^2}{4}}2^m\sqrt{\pi}m!y^{n-m}L_m^{n-m}(-y^2/2)\,,
 }
 where $L_m^n(z)$ is a generalized Laguerre polynomial,
 to compute the 2-body operator expectation value $ \sum_{i\neq j}\langle \cos(2\omega(a_i-a_j))\rangle$ in the $U(N)$ case:
\begin{align}
&\frac{N(N-1)}{N!}\sum_{\sigma\in S_N}\sum_{\mu\in S_2}(-1)^{|\mu|}\int \left(\prod_{i=1}^2da_i \frac{p_{\sigma(i)-1}(a_i)p_{\mu(\sigma(i))-1}(a_i)}{h_{\sigma(i)-1}}e^{-\frac{8\pi^2}{g_\text{YM}^2}a_i^2}\right)\frac{e^{2i\omega(a_1-a_2)}+e^{2i\omega(a_2-a_1)}}{2}\nonumber\\
&=e^{\frac{-\omega^2g_\text{YM}^2}{4\pi^2 }}\sum_{i,j=1}^N\left[L_{i-1}\left({\scriptstyle\frac{\omega^2g_\text{YM}^2}{4\pi^2}}\right)L_{j-1}\left({\scriptstyle\frac{\omega^2g_\text{YM}^2}{4\pi^2}}\right)-(-1)^{i-j}L_{i-1}^{j-i}\left({\scriptstyle\frac{\omega^2g_\text{YM}^2}{4\pi^2}}\right)L_{j-1}^{i-j}\left({\scriptstyle\frac{\omega^2g_\text{YM}^2}{4\pi^2}}\right)\right]\,.
\label{firstExp}
\end{align}
 We can then plug this into \eqref{Kfourier} and \eqref{2m2L} to get the final result for $SU(N)$ in \eqref{orthoFinal}, which recall is the same as $U(N)$. The other cases can be computed similarly, where $ \sum_{i\neq j}\langle \cos(2\omega(a_i-a_j))\rangle$ and $ \sum_{i\neq j}\langle \cos(2\omega(a_i+a_j))\rangle$ both give the same answer, for $SO(2N+1)$ and $USp(2N)$ we have an extra contribution from the 1-body operator in \eqref{2m2L}, and the $USp(2N)$ case has $g^2_\text{YM}\to g^2_\text{YM}/2$ relative to the $SO(2N+1)$ case.

\section{Closed form for $R|R$ Mellin amplitude in $SU(N)$}
\label{M_RR_App}

In this appendix we compute explicitly the double sum
\es{boxads5}{
&\Phi(s,t)= \sum_{m,n=2}^\infty \frac{c_{mn}}{(s-2m)(t-2n)}\,,
}
where the coefficients $c_{mn} = c_{nm}$ are those relevant for $SU(N)$, given in \eqref{SUNcmn}. The first remark is that this sum needs to be regularized. Different regularisations differ from each other by ambiguities of the form $\alpha+\beta(8-3s-3t)$. Below we will choose a specific regularisation and give $\alpha$ in that case. $\beta$ will not be needed for our purposes. More precisely, consider the convergent double sum
\es{boxreg}{
&\Phi_{reg}(s,t)= \sum_{m,n=2}^\infty  \left( \frac{c_{mn}}{(s-2m)(t-2n)} - \delta_{mn} \right)\,,
}
where 
\beq
\delta_{mn}= \frac{3 m n}{2 (m+n)^3}+\frac{3 m t-4 m+3 n s-4 n}{4 (m+n)^3}\,.
\eeq
Note that the subleading/second term in $\delta_{mn}$ vanishes in the symmetric combination that enters the Mellin amplitude \eqref{MellinRR}
\beq
M_{SU(N)}^{R|R}(s,t) = \Phi_{reg}(s,t)+\Phi_{reg}(t,u)+\Phi_{reg}(u,s)+C^{SU(N)}\,,
\label{mellin_summed}
\eeq
but it is necessary to regularize a single double sum, very much as what happens with the box function in flat space. In order to compute $\Phi_{reg}(s,t)$ we will follow the same strategy as in \cite{Alday:2021ajh}. The answer is expected to have the form
\begin{align}
\Phi_{reg}(s,t) ={}& R_0(s,t) \left( \psi ^{(1)}\left(2-\frac{s}{2}\right)+ \psi ^{(1)}\left(2-\frac{t}{2}\right)- \left( \psi ^{(0)}\left(2-\frac{s}{2}\right)-\psi ^{(0)}\left(2-\frac{t}{2}\right) \right)^2 \right) \nonumber\\
& + R_1(s,t) \psi ^{(0)}\left(2-\frac{s}{2}\right)+R_1(t,s) \psi ^{(0)}\left(2-\frac{t}{2}\right)+R_2(s,t)\,,
\label{AdSbox}
\end{align}
where the rational functions $R_0(s,t)$, $R_1(s,t)$, $R_2(s,t)$ are free of poles at $s,t=4,6,\cdots$. $R_0(s,t)$ is fixed by the requirement that the residues at simultaneous poles are reproduced, which simply implies that $R_0(s,t) $ is given by $c_{mn}$ upon setting $m \to s/2$ and $n \to t/2$:
\es{R0}{
R_0(s,t) = \frac{P_0(s,t)}{40(s+t-10) (s+t-8) (s+t-6) (s+t-4) (s+t-2)}\,,
}
with $P_0(s,t)$ a symmetric polynomial of total degree six
\begin{align}
P_0(s,t)={}&15 s^4 t^2+30 s^3 t^3-360 s^3 t^2+15 s^2 t^4-360 s^2 t^3+2304 s^2 t^2-70 s^4 t+1096 s^3 t\nonumber\\
&-5048 s^2 t +88 s^4-1024 s^3+3552 s^2-70 s t^4+1096 s t^3-5048 s t^2+8640 s t\nonumber\\
&-4736 s+88 t^4-1024 t^3+3552 t^2-4736 t+2048\,.
\label{P0}
\end{align}
Next, let us focus on $R_1(s,t)$. This is fixed by the requirement that at each pole $s=2m$, we get the precise $t$-dependence, namely an infinite sum over poles with the correct residues. This implies
\es{R1}{
R_1(s,t) = \frac{P_1(s,t)}{40 (s+t-10) (s+t-8) (s+t-4) (s+t-2)}\,,
}
with $P_1(s,t)$ a polynomial of total degree 5. 
\begin{align}
P_1(s,t)={}&-105 s^3 t^2-45 s^2 t^3+1200 s^2 t^2-75 s^4 t+1350 s^3 t-7720 s^2 t-15 s^5+400 s^4\nonumber\\
&-3532 s^3 +13008 s^2+250 s t^3-3980 s t^2+17200 s t-21248 s-368 t^3\nonumber\\
&+4192 t^2-13312 t+12800\,.
\end{align}
Finally, $R_2(s,t)$ can be constrained as follows. Note that because of the form of $R_0(s,t)$ and $R_1(s,t)$, spurious poles are introduced at $s+t=2,4,6,8,10$ and $R_2(s,t)$ must cancel these poles. Its polar structure together with the flat space limit (which constraints the large $s,t$ behaviour) allow to write $R_2(s,t)$ in the following form 
\es{R2}{
R_2(s,t) = \frac{P_2(s,t)}{240 (s+t-10) (s+t-8) (s+t-6) (s+t-4) (s+t-2)}+\alpha+\beta(8-3s-3t)\,,
}
where $P_2(s,t)$ is a symmetric polynomial of total degree 6. It turns out that $P_2(s,t)$ is completely fixed by requiring cancellation at the spurious poles, up to the ambiguity of the form $\alpha+\beta(8-3s-3t)$ (times the denominator), exactly as expected. For the regularisation above we obtain
\begin{align}
&P_2(s,t)=-45 s^4 t^2-630 s^3 t^2-45 s^2 t^4-630 s^2 t^3+19404 s^2 t^2-72 s^5 t+585 s^4 t \nonumber\\
&+8640 s^3 t-125604 s^2 t-27 s^6+477 s^5-1062 s^4-28908 s^3+239688 s^2-72 s t^5+585 s t^4 \nonumber\\
&+8640 s t^3-125604 s t^2+520848 s t-683424 s-27 t^6+477 t^5-1062 t^4-28908 t^3\nonumber\\
&+239688 t^2-683424 t+642816 
+\pi^2 \left(45 s^4 t^2+520 s^3 t^2+45 s^2 t^4+520 s^2 t^3-14544 s^2 t^2\right.\nonumber\\
&+54 s^5 t-400 s^4 t 
 -7056 s^3 t+89808 s^2 t+9 s^6-164 s^5-648 s^4+25984 s^3-172656 s^2\nonumber\\
&+54 s t^5-400 s t^4 
-7056 s t^3+89808 s t^2-354528 s t+458560 s+9 t^6-164 t^5-648 t^4\nonumber\\
&\left.+25984 t^3-172656 t^2 
+458560 t-419328 \right)\,,
\label{P2}
\end{align}
and for our specific normalisation $\alpha=-3\zeta(3)$. Note that $\beta$ will disappear once we consider the combination of all three channels. 
The Mellin amplitude presented here is also included in the attached \texttt{Mathematica} file.

 \section{1-loop localization integral in Mellin space}
 \label{locInt}
 
 In this Appendix we will compute the integral \eqref{Mints} for the Mellin amplitudes $M^{R|R}_{SU(N)}$ and $M^{R|R}_{orb}$ as described in Section \ref{reducedMellinAmplitude}. Our strategy is to first compute it for a general simultaneous pole in $s$ and $t$
\es{eq:I2_2poles}{
I_2 \left(\frac{1}{(s-2m)(t-2n)} \right)\,.
}
To this end, we first use the integral representation of the Gamma function
\es{}{
\Gamma(x) = \int\limits_0^\infty dt \, t^{x-1} e^{-t}\,,
}
for $\Gamma\left( \tfrac{u}{2} \right)$ and $\Gamma\left(2- \tfrac{u}{2} \right)$ in order to factorize the $s$ and $t$ integrals, and then do these integrals by summing residues
\es{}{
I_2 \left(\tfrac{1}{(s-2m)(t-2n)} \right)
= -\int\limits_0^\infty dt_1 dt_2
\frac{t_1 \Gamma (m) \Gamma (n)  \, _2F_1\left(2,m;m+1;-\frac{t_1}{t_2}\right) \,
   _2F_1\left(2,n;n+1;-\frac{t_1}{t_2}\right)}{8 e^{t_1+t_2} t_2 \Gamma (m+1) \Gamma (n+1)}\,.
}
By doing the integral for many values of $m$ and $n$, we found the general formula
\begin{align}
I_2 \left(\tfrac{1}{(s-2m)(t-2n)} \right) ={}&
\frac{1}{4} (m-1) (n-1) (m+n-1) \left(\frac{\pi ^2}{12} \left(2 H_{m+n-2}-H_{m-2}-H_{n-2}\right)- \zeta (3)\right) \nonumber\\
&-\frac{\pi ^2}{96}  (m+n) \left(m^2+m n+n^2-2 m-2 n+1\right) + r_{m,n}\,,
\label{eq:I2_st_pole}
\end{align}
where $H_n$ is the $n$-th harmonic number and $r_{m,n}$ is symmetric under exchange of $m$ and $n$ and determined by the recursion relation
\begin{align}
&r_{m,n} = -\frac{(n-1) n}{(n-3) (n-2)} r_{m,n-2}
+ \frac{2 (n-1) (m (n-1)+(n-3) n+1)}{(n-2)^2 (m+n-2)} r_{m,n-1}
\label{eq:r_mn}\\
&+\frac{m \left(6 m^2+3 m-2\right) (2 n-3)+4 n-9-6 m (m+1) (m-1)^2 (2 n-3) \left( \psi ^{(1)}(m-1) - \tfrac{\pi^2}{6} \right)}{48 (n-3) (n-2)^2 (m+n-2)}\,,
\nonumber
\end{align}
together with the initial values
\es{}{
r_{2,2} = \frac{53}{48}\,, \qquad
r_{2,3} = \frac{101}{32}\,, \qquad
r_{3,3} = \frac{223}{32}\,.
}
Next we have to deal with the double sum over $m$ and $n$. In order to improve the convergence of the sum for the $SU(N)$ case we subtract the first three terms in the large $m\sim n$ expansion of the summand \eqref{SUNcmn}. These terms are polynomial Mellin amplitudes that can be easily summed and integrated separately. We then cut off the sum of the subtracted terms as follows (using crossing symmetry of the integrand of \eqref{Mints})
\begin{align}
I_2 [M^{R|R}] \approx{}& \sum\limits_{m=2}^M \sum\limits_{n=2}^{m} \left( 3 (2-\de_{mn}) c^{SU(N)}_{mn} I_2 \left(\tfrac{1}{(s-2m)(t-2n)} \right) - b^{SU(N)}_{mn} I_2(1) -I_2(\text{subtractions}_{mn}) \right) \nonumber\\
&+\sum\limits_{m,n=2}^{\infty} I_2(\text{subtractions}_{mn})
+C^{SU(N)} I_2(1)\,.
\label{eq:I2_SGSG_A_calculation}
\end{align}
For the cutoff $M=5000$ we find agreement with the result of \cite{Chester:2019pvm} to 4 digits, as shown in \eqref{Bfix1loop}.

For the orbifold case we can reuse the $SU(N)$ result in \eqref{Bfix1loop} times $1/2$ and only need to add the terms involving $d_{mn}$. We expect the main contribution to the sum to come from the region where either $m$ or $n$ is small. In order to improve convergence in this region we expand the summand at large $n$
\es{}{
\frac{d_{mn}}{(s-2m)(t-2n)} = \text{sub}_n^{(4)} (d_{mn}) + O(n^{-5})\,.
}
The subtractions defined in this way have terms of the form
\es{}{
\frac{t^k}{s-2m}\,, \quad k=0,1,2\,,
}
for which we compute the integral \eqref{Mints}:
\es{eq:I2_t_poles}{
I_2 \left( \frac{1}{s-2m} \right) ={}& 
\frac{1}{144} \left(-6 m^3-3 m^2+2 m-2\right)
+\frac{1}{24} (m-1)^2 m (m+1) \psi ^{(1)}(m-1)\,,\\
I_2 \left( \frac{t}{s-2m} \right) ={}&
\frac{30 m^5-75 m^4+20 m^3+80 m^2-41 m+22}{720 (m-2)}\\
&-\frac{1}{24} (m-2) (m-1)^2 m (m+1) \psi ^{(1)}(m-2)\,,\\
I_2 \left( \frac{t^2}{s-2m} \right) ={}& 
\frac{-90 m^6+345 m^5-360 m^4-160 m^3+443 m^2-194 m+16}{1800 (m-2)}\\
& + \frac{1}{60} (m-2) (m-1)^2 m (m+1) (3 m-4) \psi ^{(1)}(m-2)\,.
}
The integral for $M_{orb}^{R|R}$ is then approximated by
\begin{align}
I_2 (M_{orb}^{R|R}(s,t)) \approx{}& \frac{5}{64} + \sum\limits_{m=2}^{100} \sum\limits_{n=2}^{300}  3 (2-\de_{mn}) \left( d_{mn} I_2 \left(\tfrac{1}{(s-2m)(t-2n)} \right) - I_2 (\text{sub}_n^{(4)} (d_{mn})) \right)\nonumber\\
={}& \frac{5}{64} -0.023437499999
\approx \frac{5}{64} - \frac{3}{128} \,,
\end{align}
as shown in \eqref{Bfix1loop}.
 
  \section{Lorentzian inversion}
\label{inversion}

In this Appendix we explain how to extract the CFT-data to one-loop from the double discontinuity for the $SU(N)$ and orbifold 1-loop terms. We will follow a streamlined version of the method of \cite{Alday:2017vkk}. The function that describes the CFT-data is given by
\begin{equation}
c(h,\bar h)= \int_0^1 \frac{dz}{z^2} k_{1-h}(z) \int_0^1 \frac{d \bar z}{\bar z^2} \frac{r_{\bar h}^2}{\bar h-1/2} k_{\bar h}(\bar z) \frac{dDisc[(z \bar z)(\bar z-z) \mathcal{G}(U,V)]}{4\pi^2}\,,
\end{equation}
where $h=\frac{\Delta-\ell+2}{2}$,  $\bar h=\frac{\Delta+\ell+4}{2}$ and $k_h(z)=z^h\,_2F_1(h,h,2h,z)$ and we have also introduced
\begin{equation}
r_h = \frac{\Gamma(h)^2}{\Gamma(2h-1)}\,.
\end{equation}
A given double trace operator with quantum numbers $\Delta=4+2n+\ell+\gamma$ and $\ell$ produces a pole at $h=3+n+\gamma/2$, $\bar h=4+n+\ell+\gamma/2$. For small $\gamma$ in a perturbative expansion we have
\begin{equation}
c(h,h+\ell+1)= \left\langle \frac{a}{2(3+n+\gamma/2-h)} \right\rangle_{n,\ell}\,,
\end{equation}
where the average is among all the nearly degenerate intermediate operators (degenerate at zeroth order). The inversion integral then works as follows. The OPE data is encoded in poles in the $h-$plane. These arise from the $z \to 0$ limit of the integration, and different powers of $z$ in that limit yield different twists. The integral over $\bar z$, on the other hand, is dual to $\bar h$ and provides the spin dependence for each twist.  

We will consider the inversion integral in a $1/c$ expansion and are interested in leading twist operators (of approximate twist 4). To order $1/c^2$ their contribution to the correlator will be of the form
\begin{equation}
dDisc[\mathcal{G}(U,V)] = 4\pi^2 (z \bar z)^2 \left( h_0(\bar z) \log^2 z + h_1(\bar z) \log z+ h_2(\bar z) \right) + \cdots
\end{equation}
we can then perform the integral over $z$ keeping the relevant poles in a small $z$ expansion. We obtain
\begin{equation}
c(h,\bar h)= -\frac{R_0(\bar h)}{(h-1)^3}-\frac{R_1(\bar h)}{(h-1)^2}-\frac{R_2(\bar h)}{h-1} + \cdots
\end{equation}
where we disregard all poles in $\bar h$ except the ones at $\bar h=1$ and
\bea
R_0(\bar h) &= \int_0^1 \frac{d \bar z}{\bar z^2} \frac{r_{\bar h}^2}{\bar h-1/2} k_{\bar h}(\bar z) (2 \bar z^4 h_0(\bar z))\,, \\
R_1(\bar h) &= \int_0^1 \frac{d \bar z}{\bar z^2} \frac{r_{\bar h}^2}{\bar h-1/2} k_{\bar h}(\bar z) (\bar z^4 h_1(\bar z)) \,,\\
R_2(\bar h) &= \int_0^1 \frac{d \bar z}{\bar z^2} \frac{r_{\bar h}^2}{\bar h-1/2} k_{\bar h}(\bar z) (\bar z^4 h_2(\bar z)) \,.
\eea{R012}
These residues are related to the anomalous dimensions as follows
\bea
\langle a^{(0)} \left( \gamma^{(1)}\right)^2\rangle_{\ell} &= 8 R_0(\bar h)\,, \\
\langle a^{(0)} \gamma^{(2)}+ a^{(1)} \gamma^{(1)} \rangle_{\ell} &= 4 R_1(\bar h) + 4 \partial_{\bar h}R_0(\bar h) \,.
\eea{a_to_R}

Since we are only interested in the anomalous dimension, we will only compute $R_0(\bar h)$ and $R_1(\bar h)$ for the $SU(N)$ and orbifold theories. For both theories we obtain
\begin{eqnarray}
h_0(\bar z) = 6\frac{6-6 \bar z+\bar z^2}{\bar z^6}\,.
\end{eqnarray}
$R_0(\bar h)$ can then be easily computed and gives
\begin{eqnarray}
R_0(\bar h) = \frac{24 r_{\bar h}}{(\bar h+2)(\bar h-3)}\,.
\end{eqnarray}
From this we can compute 
\begin{equation}
\langle a^{(0)} \left( \gamma^{(1)}\right)^2\rangle_{\ell} = a^{(0)} _\ell \frac{576}{(\bar h+2)^2(\bar h-3)^2}\,.
\end{equation}
Next, we want to extract the one-loop anomalous dimension. From Section \ref{1loopfrom} we find
\begin{equation}
h_1^{SU(N)}(\bar z)=\frac{18 \left({\bar z}^4+29 {\bar z}^3-204 {\bar z}^2+350 {\bar z}-175\right)}{{\bar z}^8}\,.
\end{equation}
Plugging this into the formula above we find
\begin{eqnarray}
R_1(\bar h) = \frac{36 \left({\bar h}^2-{\bar h}-55\right)}{({\bar h}-5) ({\bar h}-3) ({\bar h}+2) ({\bar h}+4)} r_{\bar h}\,.
\end{eqnarray}
Plugging this into the formula for $\langle a^{(0)} \gamma^{(2)}+ a^{(1)} \gamma^{(1)} \rangle_{\ell} $, subtracting the corresponding tree level pieces and dividing by the GFFT coefficients, we get the $c^{-2}$ values shown in \eqref{anoms}, except for spin zero where the analytic continuation only gives the contribution $\frac{287}{6}$ that neglects the contact term ambiguity.

We can now repeat the same computation for orbifold case, where we find
\begin{align}
{}&h_1^{orb}(\bar z)=
\frac{1}{262144 (1-\zb) \zb^8}
\Big(
36018675 \zb^{11}-81959850 \zb^{10}+59928960 \zb^9-14292880
   \zb^8\nonumber\\
&+347968 \zb^7-43008 \zb^6-10932224
   \zb^5-8949760 \zb^4+455073792 \zb^3-1260945408
   \zb^2\nonumber\\
&+1238630400 \zb-412876800
\Big)
+\frac{\text{arctanh}\left(\sqrt{1-\zb}\right)}{262144 (1-\zb)^{3/2}
   \zb^6}\Big(
36018675 \zb^{10}-105972300 \zb^9\nonumber\\
&+111367200 \zb^8-48484800
   \zb^7
+7083648 \zb^6-61440 \zb^5+442368
   \zb^4-10616832 \zb^3\nonumber\\
&+38535168 \zb^2-47185920
   \zb+18874368
\Big)\,.
\label{h1_SON}
\end{align}
The only difference is that now the integrals cannot be done for all $\ell$, but case by case they can be done, and one can also analytically continue to spin zero. The results are shown in \eqref{anoms}, except for spin zero we again only get the contribution $\frac{29129625 \zeta (3)+22143194}{7680}-10688 \log (2)$ that neglects the contact term ambiguity.

 \section{Mixed Correlator Bootstrap Setup}
\label{mixedCorr}

In this appendix we describe the setup for the numerical bootstrap with mixed correlators between the stress tensor
multiplet and the next lowest half-BPS multiplet.
The definitions collected from various places in the literature should be general enough to make this a useful reference for more general setups as well.
We write four-point functions of $\frac12$-BPS operators of dimension $p_i$ as
\bea
\< S_{p_1}(x_1,y_1) \cdots S_{p_4}(x_4,y_4) \>
&\equiv 
T_{p_1 p_2 p_3 p_4}
\mathcal{G}_{\{p_i\} }(z,\zb,\al,\ab)\,,\\
T_{p_1 p_2 p_3 p_4} &= g_{12}^{\frac{p_1+p_2}{2}}
g_{34}^{\frac{p_3+p_4}{2}}
\left(\frac{g_{24}}{g_{14}}\right)^{\frac{p_{2}-p_1}{2}}
\left(\frac{g_{13}}{g_{14}}\right)^{\frac{p_{3}-p_4}{2}}\,,
\eea{Eq:OriginalCorrelator}
where
\beq
g_{ij} = \frac{Y_i \cdot Y_j}{x_{ij}^2}\,, \qquad
\alpha \bar{\alpha}=\frac{1}{\sigma}\,, \qquad
(1-\alpha)(1-\bar{\alpha})=\frac{\tau}{\sigma}\,.
\eeq
The superconformal Ward identities imply that the correlator can generally be written in the form \cite{Dolan:2004iy,Caron-Huot:2018kta}
\begin{align}
\mathcal{G}_{\{p_i\} }(z, \zb, \al, \ab) &= k \chi(z, \al) \chi(\zb, \ab) + \frac{(z-\al)(z-\ab)(\zb-\al) (\zb - \ab)  }{(\al - \ab)(z - \zb)} \nonumber \\
&\times \left(- \frac{ \chi(\zb,\ab) f(z,\al)}{\al z (\zb-\ab)} + \frac{\chi(\zb,\al) f(z,\ab)}{\ab z (\zb-\al)} + \frac{\chi(z,\ab) f(\zb,\al)}{\al \zb (z-\ab)} - \frac{\chi(z,\al) f(\zb,\ab)}{\ab \zb (z-\al)} \right) \nonumber \\
&+ \frac{(z-\al) (z - \ab) (\zb- \al) (\zb - \ab)}{(\al \ab)^2} H_{\{p_i\}}(z,\zb, \al, \ab), \label{G ansatz}
\end{align}
where $k_{\{p_i\}}$ is called the unit contribution, $f_{\{p_i\}}$ is the chiral correlator and $H_{\{p_i\}}$ is the reduced correlator.
These different parts can be extracted from $\mathcal{G}_{\{p_i\} }$ as follows
\be\label{kf from G} \begin{aligned}
 k_{\{p_i\} } &= \cG_{\{p_i\} }(z, \zb, z, \zb), \\
 f_{\{p_i\} }(\zb,\ab) &= \frac{\ab \zb}{\zb - \ab} \left( \cG_{\{p_i\} }(z, \zb, z, \ab) - k_{\{p_i\}}\chi_{\{p_i\} }(\zb,\ab)  \right),
\end{aligned}\ee
and $H_{\{p_i\}}$ is obtained by subtracting the other contributions from \eqref{G ansatz}. The function $\chi_{\{p_i\}}$ is given by
\begin{equation}
\chi_{\{p_i\}}(z,\al) = \left( \frac{z}{\alpha} \right)^{\max(|p_{21}|,|p_{34}|)/2} \left( \frac{1-\al}{1-z} \right)^{\max(p_{21}+p_{34},0)/2}.
\label{Eq:ChiWard}
\end{equation}
We would like to bootstrap the contributions of long superconformal multiplets which only contribute to $H_{\{p_i\}}$. The contributions of short multiplets are completely fixed by the free theory. Combining the $k_{\{p_i\}}$ and $f_{\{p_i\}}$ from free theory with the $1 \leftrightarrow 3$ crossing equation for the correlator, we derive the following crossing equations for the reduced correlators
\bea
V^2 H_{\{2222\}}(U,V) - U^2 H_{\{2222\}}(V,U) &= \frac{(V-U) (c (U+V)+1)}{c}\,,\\
V^{5/2} H_{\{2323\}}(U,V) - U^{5/2} H_{\{2323\}}(V,U) &= \frac{(V-U) \sqrt{U V} (2 c (U+V)+3)}{2 c}\,,\\
V^{5/2} H_{\{2233\}}(U,V)-U^2 H_{\{3223\}}(V,U) &= \frac{\sqrt{V} (6 V-3 U-2 c)}{2 c}\,,\\
V^2 H_{\{3223\}}(U,V) - U^{5/2} H_{\{2233\}}(V,U) &= \frac{\sqrt{U} (3 V-6 U+2 c)}{2 c}\,,
\eea{crossing_H}
where we used the notation
\bea
H_{\{2222\}} (z,\zb,\al,\ab) &= H_{\{2222\}} (U,V)\,, &\qquad
H_{\{2323\}} (z,\zb,\al,\ab) &= \frac{1}{\sqrt{\al \ab}} H_{\{2323\}} (U,V)\,,\\
H_{\{2233\}} (z,\zb,\al,\ab) &= H_{\{2233\}} (U,V)\,, &\qquad
H_{\{3223\}} (z,\zb,\al,\ab) &= \frac{1}{\sqrt{\al \ab}} H_{\{3223\}} (U,V)\,.
\eea{eq:Hshorthand}
The reduced correlator for $\<3333\>$ is the only one with a non-trivial dependence on the R-symmetry cross-ratios. Its crossing equation is
\begin{align}
{}&V^3 H_{\{3333\}}(U,V,\sigma ,\tau )
-\tau  U^3 H_{\{3333\}}\left(V,U,\frac{\sigma }{\tau },\frac{1}{\tau }\right) =
\frac{9 (4 c-7) \left(V^2-\tau  U^2\right)}{c (4 c-3)}
+ U^4 (\sigma +\tau -1)\nonumber\\
&+U^3 (-\sigma -2 \tau +V (-2 \sigma -\tau +1)+1)+U V^3 (2 \sigma -\tau +1)+V^4 (-\sigma +\tau -1)\nonumber\\
&+V^3 (\sigma-\tau +2)
+ \frac{9}{4c} \Big( -U^3+\tau  \left(-U \left(U^2+1\right)+(U-1) V+V^3\right)
\label{crossing_H3333}\\
&+\sigma  (U-V) \left(U^2+V^2-1\right)-U V+U+V^3+V \Big).
\nonumber
\end{align}
In order to make use of the full superconformal symmetry when bootstrapping the long multiplets, we have to subtract the contribution of the short multiplets to the reduced correlator
\beq
H_{\{p_i\}}^\text{long} = H_{\{p_i\}} - H_{\{p_i\}}^\text{short} \,.
\eeq
The next subsections describe the derivation of $H_{\{p_i\}}^\text{short}$ for the mixed correlator bootstrap with external operators $S_2$ and $S_3$. This is done in several steps. First one expands the free theory correlators in terms of superconformal blocks and then takes into account multiplet recombination in order to determine which of the short multiplets remain present when transitioning from the free to the interacting theory. These steps were done for the correlators $\<2222\>$, $\<2233\>$ and $\<3333\>$ in \cite{Doobary:2015gia} and we only have to supplement these results with the corresponding ones for $\<2323\>$ and $\<3223\>$. Once the contributions of the short multiplets are determined, we need to find their contributions to the reduced correlator. The decomposition \eqref{G ansatz} and in particular the function \eqref{Eq:ChiWard} were chosen in \cite{Caron-Huot:2018kta} in such a way that the superconformal Casimir equation commutes with the decomposition, which ensures that the superconformal blocks themselves have a nice decomposition. Using this representation of the superconformal blocks it is easy to read of their contribution to the reduced correlator.

\subsection{Superconformal blocks}
\label{sec:superconformal_blocks}

The superconformal multiplets that can be exchanged in the four-point functions we will consider can be labeled by the conformal dimension $\De$, spin $\ell$ and SU(4) representation $[m,n-m,m]$ of the superconformal primary of the multiplet.
For expanding free theory correlators or using the recombination identity \eqref{recombination} it is also useful to label superconformal multiplets in a different way which handles the different types of multiplets on the same footing. Following \cite{Doobary:2015gia} we can label them by their twist $\gamma$ and a representation of $GL(2|2)$ which is labeled by a Young diagram $\ula$ of ``fat hook'' shape, meaning that with $\l_i$ the length of the $i$th row and $\l^T_j$ the height of the $j$th column of the Young diagram,
\beq
\l_i \leq 2 \,, \text{for } i \geq 3\,, \qquad
\l^T_j \leq 2 \,, \text{for } j \geq 3\,.
\eeq
One can translate between the two ways of labeling using the table below.
\begin{table}[ht]
$$
\begin{array}{|c||c|c|c|c|}
\hline
  GL(2|2) \text{ rep }\ula& \text{dim } \De & \text{spin } \ell& SU(4) \text{ rep } [m,n-m,m] &\text{multiplet} \\\hline
[0]&\gamma&0&[0,\gamma,0]&\frac12\text{-BPS} \\\hline
  [\lambda,1^{\mu}]\ (\lambda\geq 2)& \gamma{+}\lambda{-}2&\lambda{-}2&[\mu,\gamma{-}2\mu{-}2,\mu]&\text{semi-short} \\  
  \left[1^\mu\right] & \gamma&0&[\mu,\gamma{-}2\mu,\mu]& \frac14\text{-BPS} \\ \hline
{ [\lambda_1,\lambda_2,2^{\mu_2},1^{\mu_1}]\  (\lambda_2\geq 2)} & \gamma{+}\lambda_1{+}\lambda_2{-}4&\lambda_1{-}\lambda_2&[\mu_1{-}\mu_2,\gamma{-}2\mu_1-4,\mu_1{-}\mu_2]&\text{long} \\ \hline
\end{array}
$$
\caption{Translation between $\cN=4$  superconformal reps and superfields $\cO^{\gamma\ula}$.}
\label{table:sc_reps}
\end{table}
Note that for long multiplets there is a degeneracy in the description in terms of $\gamma, \ula$ since there are certain shift symmetries that leave $\De,\ell,m,n$ unchanged.

The correlator $\mathcal{G}_{\{p_i\} }(z,\zb,\al,\ab)$ can be expanded in superconformal blocks
\beq
\bar{F}^{\gamma \ula} = \left( \frac{z \zb}{\al \ab} \right)^\frac{\gamma}{2}
F^{\al \B \gamma \ula}\,, \qquad \al = \frac{\gamma+r}{2}\,, \ \B = \frac{\gamma+s}{2}\,,
\label{Fbar}
\eeq
where $r=p_{21}\equiv p_2-p_1$, $s=p_{34}$ and the definition of $F^{\al \B \gamma \ula}$ is given in \cite{Doobary:2015gia}.
The same superconformal blocks have a representation directly in terms of the components of the decomposition \eqref{G ansatz} due to \cite{Caron-Huot:2018kta}.
These have separate formulae for the different types of multiplets. We can read off from table \ref{table:sc_reps}\footnote{We checked for that the definition of the superblocks \eqref{Fbar} matches with the one below for all the cases that we are using.}
\bea
\tfrac12 \text{-BPS}:&&\quad \bar{F}^{\gamma [0]} &= \mathcal{B}_{0,\Delta}^{r,s}\,,\\
\text{semi-short}:&&\quad \bar{F}^{\gamma [\lambda,1^{\mu}]} &= \mathcal{C}_{\l-2,\mu,\gamma-\mu-2}^{r,s}\,,
\quad &&\l \geq 2\,,\\
\tfrac14\text{-BPS}:&&\quad \bar{F}^{\gamma [1^{\mu}]} &= \mathcal{B}_{\mu,\gamma-\mu}^{r,s} =  \mathcal{C}_{-1,\mu-1,\gamma-\mu-1}^{r,s}\,,\\
\text{long}:&&\quad \bar{F}^{\gamma [\lambda_1,\lambda_2,2^{\mu_2},1^{\mu_1}]} &=
\mathcal{A}_{\gamma+\l_1+\l_2-4,\l_1-\l_2,\mu_1-\mu_2,\gamma-\mu_1-\mu_2-4}^{r,s}\,, \quad &&\l_2 \geq 2\,.
\eea{eq:different_superblocks}
The block for half BPS multiplets is given by\footnote{Here we fixed several typos.}
\bea
\mathcal{B}_{0,\Delta}^{r,s} = \left\{\begin{array}{l}
\displaystyle
k=1, \qquad
f(z,\al)= \displaystyle \sum_{\substack{i=\max(|r|,|s|)\\\text{even or odd}}}^{\Delta-2}
k_{0,i}^{r,s}(z,\al) ,
\\ 
\displaystyle
H^\mathcal{B}_{\De}(z,\bar{z},\alpha, \bar{\alpha})= (z \zb)^{-2} \sum_{\substack{i=\max(|r|,|s|)\\\text{even or odd}}}^{\Delta-4}\sum_{j=0}^{(\Delta-4-i)/2}
(-1)^j G^{r,s}_{i+j+4,j} (z,\bar{z}) Z_{j,i+j}^{r,s}(\al,\ab).
\end{array}\right.
\eea{B multiplet}
The superblocks for semi-short and quarter BPS multiplets are given by
\bea
\mathcal{C}_{\ell,m,n}^{r,s} = \left\{\begin{array}{l}
\displaystyle
k=0, \qquad
f(z,\al) = (-1)^m \ k_{\ell+m+2,n-m}^{r,s}(z,\al),
\\
\displaystyle 
H^\mathcal{C}_{\ell,m,n}(z,\zb,\al,\ab) = (z \zb)^{-2} \sum_{i=1}^{m} (-1)^{i-1}\ G^{r,s}_{6+\ell+m+n-i,\ell+i}(z,\bar{z}) Z_{m-i,n-i}^{r,s}(\al,\ab)  \\
 \hspace{25mm}+ \displaystyle (z \zb)^{-2} \sum_{i=0}^{i_{\rm max}}
 (-1)^{m+i}\ G_{4+\ell+n-i,2+\ell+m+i}^{r,s}(z,\bar{z}) Z_{i,n-m-i-2}^{r,s}(\al,\ab)
\end{array}\right.
\eea{C multiplet}
with $i_{\rm max}=(n-m-2-\max(|r|,|s|))/2$.
Finally we have for long multiplets
\bea
\mathcal{A}_{\De,\ell,m,n}^{r,s} = \left\{\begin{array}{l}
\displaystyle
k=0, \qquad
f(z,\al) = 0,
\\
\displaystyle 
H^\mathcal{A}_{\De,\ell,m,n}(z,\zb,\al,\ab) = (z \zb)^{-2} G^{r,s}_{\De+4,\ell} (z,\bar{z})Z_{m,n}^{r,s}(\al,\ab)
\end{array}\right..
\eea{A multiplet}
These definitions are in terms of
the usual $SO(4,2)$ conformal blocks
\begin{align}
G_{\Delta,\ell}^{r,s} (z,\bar{z}) &= \frac{z \bar{z}}{\bar{z}-z} \left[ k_{\frac{\Delta-\ell-2}{2}}^{r,s}(z) k_{\frac{\Delta+\ell}{2}}^{r,s}(\bar{z}) - k_{\frac{\Delta+\ell}{2}}^{r,s}(z) k_{\frac{\Delta-\ell-2}{2}}^{r,s}(\bar{z}) \right], \label{Eq:G-block}
\\
k_h^{r,s}(z) &= z^h \; _{2}F_1 \left(h+ \frac{r}{2}, h + \frac{s}{2}; 2h, z \right),
\end{align}
and the $S_5$ spherical harmonics, which are given in terms of the same functions
\beq
Z_{m,n}^{r,s} (\al,\ab) = (-1)^{m} G_{-n,m}^{-r,-s}(\al,\ab)\,.
\label{eq:Z}
\eeq
We also used the notation
\bea
k_{j,m}^{r,s}(z,\al) &=  k_{1+\frac{m}2+j}^{r,s}(z) k_{-\frac{m}2}^{-r,-s}(\al)\,.
\eea{eq:Gk_atom}

\subsection{Free theory}
\label{sec:scpw_free_theory}

We begin with the correlators in the free theory
\begin{align}
\left\langle 2222\right\rangle ={}& a_1(g_{12}^2g_{34}^2+g_{13}^2g_{24}^2+g_{14}^2g_{23}^2)+a_2(g_{12}g_{23}g_{34}g_{41}+g_{13}g_{32}g_{21}g_{14}+g_{13}g_{34}g_{42}g_{21})\,,\nonumber\\
\left\langle 2233\right\rangle ={}& b_1 g_{12}^2 g_{34}^3
+b_2 \left(g_{12} g_{14} g_{23} g_{34}^2+g_{12} g_{13} g_{24}
   g_{34}^2\right)+b_3 g_{13} g_{14} g_{23} g_{24} g_{34}\,,\nonumber\\
\left\langle 3333\right\rangle ={}& c_1 \left(g_{14}^3 g_{23}^3+g_{13}^3 g_{24}^3+g_{12}^3 g_{34}^3\right)+c_2 (g_{13} g_{14}^2 g_{24} g_{23}^2+g_{12} g_{14}^2 g_{34}
   g_{23}^2 +g_{13}^2 g_{14} g_{24}^2 g_{23}\nonumber\\
&+g_{12}^2 g_{14} g_{34}^2 g_{23}+g_{12}^2 g_{13} g_{24} g_{34}^2+g_{12} g_{13}^2 g_{24}^2
   g_{34})+c_3 g_{12} g_{13} g_{14} g_{23} g_{24} g_{34}\,,
\label{eq:free_theory_correlators}
\end{align}
with
\bea
a_1 &= 1,\, a_2 = \frac{4}{N^2 -1}\,, \\
b_1 &= 1,\, b_2 = \frac{6}{N^2 -1} ,\, b_3 = \frac{12}{N^2 -1}\,, \\
c_1 &= 1,\, c_2 = \frac{9}{N^2 -1} ,\, c_3 = \frac{18(N^2-12)}{(N^2 -1)(N^2-4)}\,.
\eea{eq:abc}
The SCPW expansions of these three correlators are given by \cite{Doobary:2015gia}
\begin{align}
\left\langle 2222\right\rangle
={}& T_{2222}
\left(
1+ \sum\limits_{\l\geq 0}^\infty A^{a_2}_{2[\l]} \bar{F}^{2[\l]} + \sum\limits_{\l_1 \geq \l_2 \geq 0}^\infty A^{a_1,a_2}_{4[\l_1,\l_2]} \bar{F}^{4[\l_1,\l_2]}
\right)\,,\nonumber\\
\left\langle 2233\right\rangle
={}& T_{2233}
\left(
1+ \sum\limits_{\l\geq 0}^\infty A^{b_2}_{2[\l]} \bar{F}^{2[\l]} + \sum\limits_{\l_1 \geq \l_2 \geq 0}^\infty A^{0,b_3}_{4[\l_1,\l_2]} \bar{F}^{4[\l_1,\l_2]}
\right)\,,
\label{free_expansion}\\
\left\langle 3333\right\rangle
={}& T_{3333}
\left(
1+ \sum\limits_{\l\geq 0}^\infty A^{c_2}_{2[\l]} \bar{F}^{2[\l]} + \sum\limits_{\l_1 \geq \l_2 \geq 0}^\infty A^{c_2,c_3}_{4[\l_1,\l_2]} \bar{F}^{4[\l_1,\l_2]}
+ \hspace{-12pt}\sum\limits_{\l_1 \geq \l_2 \geq \l_3 \geq 0}^\infty \hspace{-12pt} A_{6[\l_1,\l_2,\l_3]} \bar{F}^{6[\l_1,\l_2,\l_3]}\right),\nonumber
\end{align}
where the OPE coefficients for the twist 2 and 4 multiplets only differ by their dependence on the color factors
\begin{align}
& A^a_{2[\lambda]}= \frac{1 + (-1)^\l}{2} \, \frac{2 a (\lambda!)^2}{(2 \lambda)!}\,,\\
&\notag A^{a,b}_{4[\lambda_{1},\lambda_{2}]}= \frac{1 + (-1)^{\lambda_{1}-\lambda_{2}}}{2} \, \frac{\lambda _1! \left(\lambda _1+1\right)! \left(\lambda _2!\right){}^2 \left(a \left(\lambda _1-\lambda _2+1\right) \left(\lambda _1+\lambda _2+2\right)+b (-1)^{\lambda_2}\right)}{\left(2 \lambda _2\right)! \left(2\lambda _1+1\right)!}\,,
\end{align}
and the coefficients for the twist 6 operators in the $\<3333\>$ correlator are given by
\begin{align}\label{eq:3333cof}
\notag A_{6[\lambda_1,\lambda_2]}={}& m^+_{\lambda_1,\lambda_2} \frac{1}{2} \Big(c_1\left(\lambda _1+2\right) \left(\lambda _1+3\right) \left(\lambda _1-\lambda _2+1\right) \left(\lambda _2+1\right) \left(\lambda _2+2\right) \left(\lambda
   _1+\lambda _2+4\right)\\
&\notag+4 c_2
    \left(\left((-1)^{\lambda _2}+1\right) \lambda _1 \left(\lambda _1+5\right)+8 (-1)^{\lambda _2}+\left((-1)^{\lambda _2}-1\right) \lambda _2 \left(\lambda
   _2+3\right)+4\right)\Big)\,,\\
\notag A_{6[\lambda_{1},\lambda_{2},1]}={}&m^-_{\lambda_1,\lambda_2}\frac{1}{4} \Big(c_1 \left(\lambda _1+1\right) \left(\lambda _1+4\right) \left(\lambda _1-\lambda _2+1\right) \lambda _2 \left(\lambda _2+3\right) \left(\lambda _1+\lambda _2+4\right)\\
&+4 c_2\left((-1)^{\lambda _2}-1\right) \left(\lambda _1-\lambda _2+1\right) \left(\lambda _1+\lambda _2+4\right)\Big)\,,\\
\notag A_{6[\lambda_1,\lambda_2,2]}={}&m^+_{\lambda_1,\lambda_2}\frac{1}{12} \Big(c_1 \lambda _1 \left(\lambda _1+5\right) \left(\lambda _1-\lambda _2+1\right) \left(\lambda _2-1\right) \left(\lambda _2+4\right) \left(\lambda _1+\lambda _2+4\right)\\&\notag+4 c_2\left(\left((-1)^{\lambda _2}+1\right) \lambda _1 \left(\lambda _1+5\right)+\left((-1)^{\lambda _2}-1\right) \left(\lambda _2-1\right) \left(\lambda
   _2+4\right)\right)\Big)\,,
\end{align}
with
\begin{align}m^{\pm}_{\lambda_{1},\lambda_{2}}=\frac{1 \pm (-1)^{\lambda_{1}-\lambda_{2}}}{2} \, \frac{\left(\lambda _1+2\right)!{}^2 \left(\lambda _2+1\right)!{}^2}{\left(2\lambda_1+4\right)! \left(2\lambda_2+2\right)! }.\end{align}

The correlators $\<2323\>$ and $\<3223\>$ are related to $\<2233\>$ in \eqref{eq:free_theory_correlators} by crossing and we obtain their SCPW expansion by using 
the elegant method from \cite{Doobary:2015gia} where one expands both the superconformal blocks as well as the correlator in terms of super Schur polynomials and equates the coefficients.
We find the expansion
\beq
\left\langle 2323\right\rangle
=T_{2323}
\left(
\sum\limits_{\l\geq 0}^\infty A_{3[\l]} \bar{F}^{3[\l]} + \sum\limits_{\l_1 \geq \l_2 \geq 0}^\infty A_{5[\l_1,\l_2]} \bar{F}^{5[\l_1,\l_2]}
\right)\,,
\label{2323_free_expansion}
\eeq
with the coefficients
\begin{align}
A_{3[\l]} ={}& \frac{\pi  2^{-4 \lambda -3} \left(b_2 (-1)^{\lambda } (\lambda +1)+b_3\right) \Gamma (2 \lambda +3)}{\Gamma
   \left(\lambda +\frac{3}{2}\right)^2}\,,\nonumber\\
A_{5[\l_1,\l_2]} ={}& \frac{\pi  \Gamma \left(\lambda
   _1+3\right) \Gamma \left(\lambda _2+2\right)}{2^{2 (\lambda_1+\lambda_2)+5} \Gamma \left(\lambda _1+\frac{5}{2}\right) \Gamma \left(\lambda
   _2+\frac{3}{2}\right)}
\Big(
2 b_2 \left((-1)^{\lambda _1}
   \left(\lambda _1+2\right)+(-1)^{\lambda _2} \left(\lambda _2+1\right)\right)\nonumber\\
&+b_1 (-1)^{\lambda _1-\lambda _2} \left(\lambda _1+2\right) \left(\lambda _1-\lambda
   _2+1\right) \left(\lambda _2+1\right) \left(\lambda _1+\lambda _2+3\right)
\Big)\,.
\label{OPE_2323}
\end{align}
Similarly we find the following expansion for the correlator $\<3223\>$
\beq
\left\langle 3223\right\rangle
=T_{3223}
\left( \sum\limits_{\l\geq 0}^\infty A_{3[\l]} \bar{F}^{3[\l]} + \sum\limits_{\l_1 \geq \l_2 \geq 0}^\infty A_{5[\l_1,\l_2]} \bar{F}^{5[\l_1,\l_2]}
\right)\,.
\eeq
The OPE coefficients are in this case
\begin{align}
A_{3[\l]} ={}& 
\frac{\pi  2^{-4 \lambda -3} \left(b_2 (\lambda +1)+b_3 (-1)^{\lambda }\right) \Gamma (2 \lambda +3)}{\Gamma
   \left(\lambda +\frac{3}{2}\right)^2}
\,,\nonumber\\
A_{5[\l_1,\l_2]} ={}&
\frac{\pi  \Gamma \left(\lambda
   _1+3\right) \Gamma \left(\lambda _2+2\right)}{ 2^{2 (\lambda _1+ \lambda_2)+5} \Gamma \left(\lambda _1+\frac{5}{2}\right) \Gamma
   \left(\lambda _2+\frac{3}{2}\right)}
 \Big(2 b_2 \left( (-1)^{\lambda _1} \left(\lambda_2+1\right)+ (-1)^{\lambda _2}\left(\lambda _1+2\right) \right)\nonumber\\
& +b_1 \left(\lambda _1+2\right) \left(\lambda _1-\lambda _2+1\right)
   \left(\lambda _2+1\right) \left(\lambda _1+\lambda _2+3\right)\Big) 
\,.
\label{OPE_3232}
\end{align}

\subsection{Short sector in the interacting theory}

In order to identify the short sector in the interacting theory it is necessary to resolve an ambiguity between short multiplets and long multiplets at the unitarity bound which is due to the recombination identity
\begin{align}\bar{F}_{\text{long}}^{\gamma[\lambda+1,1^{\nu+1}]}\equiv\text{lim}_{\rho\rightarrow 1}\bar{F}^{\gamma[\lambda+\rho,\rho,1^\nu]}=
\bar{F}^{\gamma-2[\lambda+2,1^{\nu}]}+\bar{F}^{\gamma[\lambda+1,1^{\nu+1}]}\,.
\label{recombination}
\end{align}
Here the LHS is the analytic continuation of a long multiplet with $\rho=2,3,\ldots$ to the unitarity bound where it becomes equal to the sum of two short multiplets.
This can be used to replace short multiplets by multiplets of higher twist $\gamma$.

For the correlators $\<2222\>$, $\<2233\>$ and $\<3333\>$ the short sectors in the interacting theory were determined already in \cite{Doobary:2015gia} and we repeat them here for convenience, supplemented with the results for $\<2323\>$ and $\<3223\>$.
In order to resolve this ambiguity we follow section 2.4 of \cite{Aprile:2019rep}.
There it is argued that in a large $c$ expansion the OPE coefficients of protected multiplets with twist $\gamma < \min(p_1+p_2,p_3+p_4)$ have to be $O(c^{-2})$. However, in the free theory correlators above, all twist 2 and twist 3 semi-short multiplets have OPE coefficients with contributions only at $O(1)$ and $O(c^{-1})$, so we have to use \eqref{recombination} to remove them completely and replace them by twist 4 and  5 multiplets.
Another way to argue is that there are no double trace semi-short operators of the form $S_q \partial^\ell S_{\tilde q}$ (with $S_q$ and $S_{\tilde q}$ half BPS operators)  of twist $q + \tilde q = 2$ or $3$ that could be protected.
For the correlators involving $S_2$ we thus have
\bea
\mathcal{G}_{\{2222\} / \{2233\}}^\text{short} ={}& 
1+ A_{2[0]}\bar{F}^{2[0]}+ \sum_{\lambda \geq 0}^{\infty} A_{4[\lambda]}\bar{F}^{4[\lambda]}+\sum_{\lambda \geq 1}^{\infty}A'_{4[\lambda,1]}\bar{F}^{4[\lambda,1]}\,,\\
\mathcal{G}_{\{2323\} / \{3223\}}^\text{short} ={}& A_{3[0]} \bar{F}^{3[0]}
 + \sum\limits_{\l \geq 0}^\infty A_{5[\l]} \bar{F}^{5[\l]}
 + \sum\limits_{\l \geq 1}^\infty A'_{5[\l,1]} \bar{F}^{5[\l,1]}\,,
\eea{Gshort}
where
\bea
A'_{4[\lambda,1]}&=A_{4[\lambda,1]}-A_{2[\lambda+1]}\,,\\
A'_{5[\l,1]} &= A_{5[\l,1]} - A_{3[\l+1]}\,,
\eea{Aprime}
and it is understood that the different OPE coefficients for each correlator from section \ref{sec:scpw_free_theory} are used.

For the correlator $\<3333\>$ the same argument works for the twist 2 semi-short multiplets, however the OPE coefficients of the twist 4 multiplets have terms at $\cO(c^{-2})$, so they can be present in the interacting theory. Since these operators are non-degenerate, these OPE coefficients are however determined by OPE coefficients that appear in other correlators \cite{Doobary:2015gia}
\begin{align}&\tilde{A}_{4[\lambda]}=\frac{\left(A_{4[\lambda]}^{ 2233}\right)^2}{A_{4[\lambda]}^{ 2222}}= \frac{1+(-1)^{\l}}{2} \, \frac{144 \lambda ! (\lambda +1)!}{(N^2-1) (2 \lambda +1)! \left(-\lambda  (\lambda +3)+(\lambda
   +1) (\lambda +2) N^2+2\right)},\nonumber\\
& \tilde{A}_{4[\lambda,1]}=\frac{\left(A_{4[\lambda,1]}^{'2233}\right)^2}{A_{4[\lambda,1]}^{' 2222}}= \frac{1-(-1)^{\l}}{2} \, \frac{576  ((\lambda +1)!)^2}{(N^2-1) (2 \lambda +2)! \left(\lambda  (\lambda +3)
   \left(N^2-1\right)-12\right)}.
\label{A4tilde}
\end{align}
With these coefficients for the twist 4 multiplets, the short contributions are completely determined\footnote{We corrected the coefficient for $\bar{F}^{6[1,1]}$ in \cite{Doobary:2015gia}.}
\bea
\mathcal{G}_{\{3333\}}^\text{short} = 
{}&c_1+A_{2[0]}\bar{F}^{2[0]}+A_{4[0]}\bar{F}^{4[0]}+\sum_{\lambda\geq 2}\tilde{A}_{4[\lambda]}\bar{F}^{4[\lambda]}+\sum_{\lambda\geq 1}\tilde{A}_{4[\lambda,1]}\bar{F}^{4[\lambda,1]}\\[10pt]
&+\sum_{\lambda\geq 0}A_{6[\lambda]}\bar{F}^{6[\lambda]}+\sum_{\lambda\geq 1}A'_{6[\lambda,1]}\bar{F}^{6[\lambda,1]}+\sum_{\lambda\geq 2}A'_{6[\lambda,1,1]}\bar{F}^{6[\lambda,1,1]}\,,
\eea{Gshort3333}
with
\bea A'_{6[\lambda,1]}&=A_{6[\lambda,1]}-A_{4[\lambda+1]}+\tilde{A}_{4[\lambda+1]}, \\
A'_{6[\lambda,1,1]}&=A_{6[\lambda,1,1]}-A_{4[\lambda+1,1]}+A_{2[\lambda+2]}+\tilde{A}_{4[\lambda+1,1]}.
\eea{A6p}

\subsection{Short contributions to reduced correlators}

In order to determine the protected parts of the reduced correlators, we simply 
read of the contributions of the superconformal blocks to $H_{\{p_i\}}$ from the definitions in section \ref{sec:superconformal_blocks}
\bea
H_{\{2222\} / \{2233\}}^\text{short} ={}&
A_{4[0]}H^\mathcal{B}_{4} + \sum_{\lambda \geq 2}^{\infty} A_{4[\lambda]}H^\mathcal{C}_{\l-2,0,2}+\sum_{\lambda \geq 1}^{\infty}A'_{4[\lambda,1]}H^\mathcal{C}_{\l-2,1,1}\,,\\
H_{\{2323\} / \{3223\}}^\text{short} ={}& A_{5[0]} H^\mathcal{B}_{5}
 + \sum\limits_{\l \geq 1}^\infty A_{5[\l]} H^\mathcal{C}_{\l-2,0,3}
 + \sum\limits_{\l \geq 1}^\infty A'_{5[\l,1]} H^\mathcal{C}_{\l-2,1,2}\,,\\
H_{\{3333\}}^\text{short} ={}& 
A_{4[0]}H^\mathcal{B}_{4} + A_{6[0]}H^\mathcal{B}_{6} +\sum_{\lambda\geq 2}\tilde{A}_{4[\lambda]}H^\mathcal{C}_{\l-2,0,2}+\sum_{\lambda\geq 1}\tilde{A}_{4[\lambda,1]}H^\mathcal{C}_{\l-2,1,1}\\
&+\sum_{\lambda\geq 2}A_{6[\lambda]}H^\mathcal{C}_{\l-2,0,4}+\sum_{\lambda\geq 1}A'_{6[\lambda,1]}H^\mathcal{C}_{\l-2,1,3}+\sum_{\lambda\geq 2}A'_{6[\lambda,1,1]}H^\mathcal{C}_{\l-2,2,2}\,.
\eea{eq:Hshort_H}
Now it is a matter of inserting further definitions to obtain the following expressions in terms of standard conformal blocks that can be used as input for the numerical bootstrap
\bea
{}&U^{2} H_{\{2222\}}^\text{short}(U,V) = \left(\frac{1}{c}+2\right) G^{0,0}_{4,0}(z,\zb)\\
&+ \sum\limits_{\ell=0}^\infty
\frac{\left((-1)^\ell+1\right) (c (\ell+3) (\ell+4)+1) \Gamma (\ell+3) \Gamma (\ell+4) }{2 c \Gamma (2 \ell+6)} G^{0,0}_{\ell+6,\ell+2}(z,\zb)\\
&+ \sum\limits_{\ell=-1}^\infty
-\frac{\sqrt{\pi } 2^{-2 \ell-7} \left((-1)^\ell-1\right) (c (\ell+2) (\ell+5)-3) \Gamma (\ell+4) }{c \Gamma
   \left(\ell+\frac{7}{2}\right)} G^{0,0}_{\ell+7,\ell+1}(z,\zb)\,,
\eea{Hshort2222}
\bea
{}&U^{2} H_{\{2233\}}^\text{short}(U,V) =
\frac{3}{c} G^{0,0}_{4,0}(z,\zb)
+\sum\limits_{\ell=0}^\infty
\frac{3 \left((-1)^\ell+1\right) \Gamma (\ell+3) \Gamma (\ell+4) }{2 c \Gamma (2 \ell+6)} G^{0,0}_{\ell+6,\ell+2}(z,\zb)\\
&+ \sum\limits_{\ell=-1}^\infty
\frac{3 \sqrt{\pi } 4^{-\ell-3} \left((-1)^\ell-1\right) \Gamma (\ell+4)}{c \Gamma
   \left(\ell+\frac{7}{2}\right)} G^{0,0}_{\ell+7,\ell+1}(z,\zb)\,,
\eea{Hshort2233}
\bea
{}&U^{2} H_{\{2323\}}^\text{short}(U,V) =
\left(\frac{3}{2 c}+1\right) G^{1,-1}_{5,0}(z,\zb)\\
&+ \sum\limits_{\ell=-1}^\infty
\frac{\sqrt{\pi } 4^{-\ell-4} \left((-1)^\ell (\ell+4) (c (\ell+3) (\ell+5)+3)+3\right) \Gamma (\ell+5) }{c \Gamma
   \left(\ell+\frac{9}{2}\right)} G^{1,-1}_{\ell+7,\ell+2}(z,\zb)\\
&+ \sum\limits_{\ell=-1}^\infty
\frac{\sqrt{\pi } 2^{-2 \ell-7} \left((-1)^{\ell+1} (\ell+4) (c (\ell+2) (\ell+6)-6)-12\right) \Gamma (\ell+5) }{3 c \Gamma \left(\ell+\frac{9}{2}\right)} G^{1,-1}_{\ell+8,\ell+1}(z,\zb)\,,
\eea{Hshort2323}
\bea
{}&U^{2} H_{\{3223\}}^\text{short}(U,V) =
\left(\frac{3}{2 c}+1\right) G^{-1,-1}_{5,0}(z,\zb)\\
&+ \sum\limits_{\ell=-1}^\infty
\frac{\sqrt{\pi } 4^{-\ell-4} \left(c (\ell+3) (\ell+4) (\ell+5)+3 \left(\ell+(-1)^\ell+4\right)\right) \Gamma (\ell+5)
   }{c \Gamma \left(\ell+\frac{9}{2}\right)} G^{-1,-1}_{\ell+7,\ell+2}(z,\zb)\\
&+ \sum\limits_{\ell=-1}^\infty
\frac{\sqrt{\pi } 2^{-2 \ell-7} \left((\ell+4) (c (\ell+2) (\ell+6)-6)+12 (-1)^\ell\right) \Gamma (\ell+5) }{3 c \Gamma \left(\ell+\frac{9}{2}\right)}G^{-1,-1}_{\ell+8,\ell+1}(z,\zb)\,.
\eea{Hshort3223}
and
\beq 
H_{\{3333\}}^\text{short}(U,V,\sigma,\tau) 
= Z_{0,0} H_{\{3333\}}^{\text{short},0}(U,V)
+2 Z_{0,2} H_{\{3333\}}^{\text{short},1}(U,V)  
- Z_{1,1} H_{\{3333\}}^{\text{short},2}(U,V)\,,
\eeq
depends on the spherical harmonics \eqref{eq:Z} for the three R-symmetry representations
\beq
Z_{0,0} = 1\,, \quad
2 Z_{0,2} = \sigma+\tau-\frac13\,, \quad
-Z_{1,1} = \sigma - \tau\,.
\eeq
The three contributions are given by
\begin{align}
{}& U^2 H_{\{3333\}}^{\text{short},0}(U,V) =
\frac{(16 c (c+6)-153) }{2 c (4 c-3)} G^{0,0}_{4,0}(z,\zb)\nonumber\\
&+ \sum\limits_{\ell=0}^\infty
\frac{9 \left((-1)^\ell+1\right) \Gamma (\ell+3) \Gamma (\ell+4)
   }{2 c (c (\ell+3) (\ell+4)+1)   \Gamma (2 \ell+6)} G^{0,0}_{\ell+6,\ell+2}(z,\zb)\nonumber\\
&- \sum\limits_{\ell=-1}^\infty
\frac{18 \left((-1)^\ell-1\right) \Gamma (\ell+4)^2 }{c (c (\ell+2) (\ell+5)-3) \Gamma (2 \ell+7)}
G^{0,0}_{\ell+7,\ell+1}(z,\zb)\nonumber\\
&+ \sum\limits_{\ell=-1}^\infty
\frac{3 \sqrt{\pi } 2^{-2 \ell-4} \left((-1)^\ell-1\right) \Gamma (\ell+5)}{\Gamma
   \left(\ell+\frac{9}{2}\right)} 
\bigg(\frac{1}{4 c-3}-\frac{3 (\ell+4) (\ell+5)}{16 (c (\ell+4) (\ell+5)+1)}-\frac{3 \ell (\ell+9)+76}{64 c}\nonumber\\
&\qquad \quad +\frac{1}{192} (\ell+2) (\ell+4) (\ell+7) (\ell+5)\bigg)  G^{0,0}_{\ell+7,\ell+3}(z,\zb)\nonumber\\
&+ \sum\limits_{\ell=0}^\infty
\frac{3 \sqrt{\pi } 2^{-2 \ell-5} \left((-1)^\ell+1\right) \Gamma (\ell+5)}{\Gamma \left(\ell+\frac{9}{2}\right)}
 \bigg(\frac{1}{4 c-3}-\frac{(\ell+3) (\ell+6)}{4 (c (\ell+3) (\ell+6)-3)}+\frac{(\ell+2) (\ell+7)}{16
   c}\nonumber\\
&\qquad \quad -\frac{1}{288} (\ell+2) (\ell+3) (\ell+7) (\ell+6)\bigg)  G^{0,0}_{\ell+8,\ell+2}(z,\zb)\,,
\label{Hshort0_3333}
\end{align}
\begin{align}
{}& U^2 H_{\{3333\}}^{\text{short},1}(U,V) =
\left(1 +\frac{9}{4c}\right) G^{0,0}_{6,0}(z,\zb)\nonumber\\
&+ \sum\limits_{\ell=0}^\infty
\frac{\left((-1)^\ell+1\right) (\ell+4) (\ell+5) (c (\ell+3) (\ell+6)+9) \Gamma (\ell+5)^2 }{8 c \Gamma (2 \ell+9)} G^{0,0}_{\ell+8,\ell+2}(z,\zb)\nonumber\\
&- \sum\limits_{\ell=-1}^\infty
\frac{3 \sqrt{\pi } 2^{-2 \ell-5} \left((-1)^\ell-1\right) \Gamma (\ell+5) }{\Gamma \left(\ell+\frac{9}{2}\right)}
   \bigg(\frac{1}{4
   c-3}-\frac{3 (\ell+4) (\ell+5)}{16 (c (\ell+4)
   (\ell+5)+1)}-\frac{3 \ell (\ell+9)+76}{64 c}\nonumber\\
&\qquad \quad +\frac{1}{192} (\ell+2) (\ell+4) (\ell+7) (\ell+5)\bigg) G^{0,0}_{\ell+9,\ell+1}(z,\zb)\,,
\label{Hshort1_3333}
\end{align}
\begin{align}
{}& U^2 H_{\{3333\}}^{\text{short},2}(U,V) =
\left(2+\frac{9}{2 c}\right) G^{0,0}_{5,1}(z,\zb)\nonumber\\
&+ \sum\limits_{\ell=0}^\infty
\frac{\left((-1)^\ell+1\right) (\ell+4) (\ell+5) (c (\ell+3) (\ell+6)+9)
   \Gamma (\ell+5)^2 }{4 c \Gamma (2 \ell+9)} G^{0,0}_{\ell+7,\ell+3}(z,\zb)\nonumber\\
&+ \sum\limits_{\ell=0}^\infty
\frac{3 \sqrt{\pi } 2^{-2 \ell-5} \left((-1)^\ell+1\right) \Gamma (\ell+5) }{\Gamma
   \left(\ell+\frac{9}{2}\right)}
   \bigg(\frac{1}{4 c-3}-\frac{(\ell+3) (\ell+6)}{4 (c (\ell+3)
   (\ell+6)-3)}+\frac{(\ell+2) (\ell+7)}{16 c}\nonumber\\
&\qquad \quad -\frac{1}{288} (\ell+2) (\ell+3) (\ell+7) (\ell+6)\bigg)
G^{0,0}_{\ell+9,\ell+1}(z,\zb)\,.
\label{Hshort2_3333}
\end{align}
For all cases except $\<3333\>$ it is straightforward to perform the sums using the Euler type integral representation for the hypergeometric function. For $\<2222\>$ this leads to the known 
expression in \cite{Beem:2016wfs}.

\bibliographystyle{JHEP}
\bibliography{4ddraft}

\end{document}